\documentclass[review]{elsarticle}
\usepackage[bookmarks=false]{hyperref}
\usepackage{graphicx,amsfonts,bm,epsfig,color,latexsym,url}
\usepackage{amsfonts}
\usepackage{amssymb}
\usepackage{epstopdf}
\usepackage{geometry}
\usepackage{subcaption}

\begin{document}
	\newcommand{\bd}{\begin{document}}
		\newcommand{\ed}{\end{document}}
	\newcommand{\bc}{\begin{center}}
		\newcommand{\ec}{\end{center}}
	\newcommand{\bfr}{\begin{flushright}}
		\newcommand{\efr}{\end{flushright}}
	\newcommand{\lt}{\left}
	\newcommand{\rt}{\right}
	\newcommand{\vs}{\vspace}
	\newcommand{\hs}{\hspace}
	
	\newcommand{\beq}{\begin{equation}}
		\newcommand{\eeq}{\end{equation}}
	\newcommand{\bea}{\begin{eqnarray}}
		\newcommand{\eea}{\end{eqnarray}}
	\newcommand{\bes}{\begin{subequations}}
		\newcommand{\ees}{\end{subequations}}
	
	\newtheorem{thrm}{Theorem}[section]
	\newtheorem{note}{Note}[section]
	\newtheorem{case}{Case}[section]
	\newtheorem{dfn}{Definition}[section]
	\newtheorem{ex}{Example}[section]
	\newtheorem{subex}{Example}[subsection]
	\newtheorem{cl}{Corrolary}[section]
	\newtheorem{propo}{Proposition}[section]
	
	\newcommand{\lb}{\linebreak}
	\newcommand{\pb}{\pagebreak}
	\newcommand{\mb}{\makebox}
	\newcommand{\fb}{\framebox}
	\newcommand{\mc}{\multicolumn}
	\newcommand{\ben}{\begin{enumerate}}
		\newcommand{\een}{\end{enumerate}}
	\newcommand{\bit}{\begin{itemize}}
		\newcommand{\eit}{\end{itemize}}
	\newcommand{\un}{\underline}
	\newcommand{\lefq}{\lefteqn}
	\newcommand{\ba}{\begin{array}}
		\newcommand{\ea}{\end{array}}
	\newcommand{\beqa}{\begin{eqnarray}}
		\newcommand{\eeqa}{\end{eqnarray}}
	\newcommand{\beqas}{\begin{eqnarray*}}
		\newcommand{\eeqas}{\end{eqnarray*}}
	\newcommand{\bfg}{\begin{figure}}
		\newcommand{\efg}{\end{figure}}
	\newcommand{\bds}{\begin{displaymath}}
		\newcommand{\eds}{\end{displaymath}}
	\newcommand{\btb}{\begin{tabbing}}
		\newcommand{\etb}{\end{tabbing}}
	\newcommand{\para}{\parallel}
	\newcommand{\pad}{\partial}
	\newcommand{\nn}{\nonumber}
	\newcommand{\la}{\leftarrow}
	\newcommand{\ra}{\rightarrow}
	\newcommand{\lgla}{\longleftarrow}
	\newcommand{\lgra}{\longrightarrow}
	\newcommand{\La}{\Leftarrow}
	\newcommand{\Ra}{\Rightarrow}
	\newcommand{\Lra}{\Leftrightarrow}
	\newcommand{\Lgla}{\Longleftarrow}
	\newcommand{\Lgra}{\Longrightarrow}
	\newcommand{\lan}{\langle}
	\newcommand{\ran}{\rangle}
	\renewcommand{\a}{\alpha}
	\renewcommand{\b}{\beta}
	\newcommand{\g}{\gamma}
	\newcommand{\G}{\Gamma}
	\renewcommand{\d}{\delta}
	\newcommand{\eps}{\epsilon}
	\newcommand{\Th}{\Theta}
	\newcommand{\s}{\sigma}
	\newcommand{\lam}{\lambda}
	\newcommand{\D}{\Delta}
	\newcommand{\vare}{\varepsilon}
	\newcommand{\pr}{\prime}
	\newcommand{\ro}{\rho}
	\newcommand{\nab}{\nabla}
	\newcommand{\m}{\mu}
	\newcommand{\n}{\nu}
	\newcommand{\Sg}{\Sigma}
	\newcommand{\p}{\pi}
	\newcommand{\R}{I\!\!R}
	\newcommand{\om}{\omega}
	\newcommand{\Om}{\Omega}
	\newcommand{\ze}{\zeta}
	\newcommand{\vart}{\vartheta}
	\newcommand{\tri}{\triangle}
	\newcommand{\f}{\frac}
	\newcommand{\ds}{\displaystyle}
	\newcommand{\iny}{\infty}
	\newcommand{\pro}{\propto}
	\newcommand{\np}{\newpage}
	\begin{frontmatter}
		\title{Stability analysis of multiple solutions of three wave interaction with group velocity dispersion and wave number mismatch}
		\author[address1]{Niladri Ghosh}
		\ead{niladri.02mgf@gmail.com}
		\author[address1]{Amiya Das}
		\ead{amiya620@gmail.com}
		\author[address2]{Debraj Nath\corref{mycorrespondingauthor}}
		\cortext[mycorrespondingauthor]{Corresponding author}
		\ead{debrajn@gmail.com}
		\address[address1]{Department of Mathematics, University of Kalyani, Kalyani - 741235, India.}
		\address[address2]{Department of Mathematics, Vivekananda College, Kolkata - 700063, India.}
		\date{\today}
		%\maketitle
		\begin{abstract}
			This paper explores the analytical approach for obtaining the multiple solutions of three-wave interacting system in $(1+1)$ dimensions. We present a novel approach by expressing the wave solutions in terms of Jacobi elliptic functions and delve into specific cases involving hyperbolic functions. Additionally, the paper focuses on analysing the linear stability of two kinds of solutions: (a) periodic and (b) one or two-hump bright solitons due to group velocity and group velocity dispersion. For linear stability, we solve the eigenvalue problem by Fourier collocation method, where Fourier coefficients are defined analytically and compared numerically. On the other hand, we check the linear stability by direct numerical simulations with Pseudospectral method along special derivatives $(t)$ and 4th order Runge-Kutta method in temporal direction $(z)$. Then it is confirmed by Crank-Nicholson finite difference method. Furthermore, we introduce a special case known as constant magnitude wave solution and examine its modulational instability in presence of group velocity dispersion. In addition, the influence of group velocities and wave vector mismatch are investigated. 
		\end{abstract}
		
	\end{frontmatter}
	\section{Introduction}
	Three wave interaction (TWI) plays a crucial role in nonlinear physics and widely applicable across various disciplines. This theory processes the parametric amplification, where energy is transferred from an external excitation (referred to as the pump wave) to a pair of daughter waves known as the signal and idler waves. The temporal coordinate $t$ can be treated as a spatial coordinate due to well-established space-time analogy in the governing three-wave interaction equations. In such cases, soliton solutions exhibit an intriguing characteristic known as beam trapping, which has significant attention in both theoretical and experimental research \cite{k96.8}. Typically, the soliton propagation effect of wave is well known in presence of cubic ($\chi^{(3)}$) nonlinearity. However, in previous literature, a range of phenomena utilizing quadratic ($\chi^{(2)}$) nonlinearity, which arises due to $\chi^{(3)}$ interaction, has been discussed in ref. \cite{k96.8}. The lowest order nonlinearity ($\chi^{(2)}$) has important role in  physics and as well as in diverse fields like hydrodynamics \cite{a.p.and.p.a3}, plasma physics \cite{a.p.and.p.a5}, fluid dynamics \cite{fluid1}, nonlinear acoustics \cite{a.p.and.p.a4}, nonlinear optics \cite{a.p.and.p.a9} and matter waves \cite{a.p.and.p.a11.2}. It is now understood that nonlinear materials possessing second-order ($\chi^{(2)}$) nonlinearities are not limited to harmonic and parametric generation. It has applications in the theory of optical switching. Experimental work has been observed an finding a new form of soliton in $\chi^{(2)}$ media, demonstrating its utility in second-harmonic and sum-frequency generation \cite{twnri1st4,twnri2nd}. The analytical soliton solutions of TWI in absence of group velocity dispersion have been defined in \cite{twnri1st4} for second-harmonic pulse propagation. Theoretical exploration of solitons of TWI in a media with quadratic nonlinearity have been solved in \cite{twri1st, twri1st2}. The analytical solutions are highly effective and valuable for investigating TWI processes and it is important for understanding the process of three wave soliton interaction \cite{twnri2nd2}. The soliton of TWI cannot be demonstrated by simple two-pole three wave soliton solutions.
	
	Resonant interactions are of great importance as they facilitate the transfer of power (energy) between a fundamental wave and its corresponding second harmonic, as well as among three plane electromagnetic waves that satisfy the energy relationship $\omega_3=\omega_1+\omega_2$ and momentum relationship for wave vectors such that $\mathbf{k_3}=\mathbf{k_1}+\mathbf{k_2}+\Delta \mathbf{k}$, where $|\Delta \mathbf{k}|\ll |\mathbf{k_3}|$ (referred to as wave vector mismatch)  \cite{k97.7,a.p.and.p.a1}. The three-wave resonant interaction (TWRI) occurs when waves with different frequencies combine in a weakly nonlinear dispersive medium and satisfying the aforementioned relations  \cite{a.p.and.p.a12,a.p.and.p.a13.3,a.p.and.p.a13.4,a.p.and.p.a14,a.p.and.p.a15.6,a.p.and.p.a15.7,a.p.and.p.a15.8,a.p.and.p.a15.9,a.p.and.p.a16.2,a.p.and.p.a16.3,a.p.and.p.a16.4,a.p.and.p.a,cnsns}. A particular type of nondegenerate TWRI is termed as type II second-harmonic generation (see ref \cite{kref96}) in nonlinear optics. It is to be mentioned here that the degenerate case is studied frequently in the literature \cite{deg}. When two identical waves having fundamental frequency yield a second-harmonic wave, they are termed as type I second-harmonic generation \cite{k96.8,twri1st,a.p.and.p.a16.2,kref97}. In absence of group velocity dispersion, the envelope of first order three wave resonant interaction has been investigated in the refs.  \cite{twri1st,twri1st2,a.p.and.p.a14,twri1st5,twri1st6,twri1st7,twri1st8,twri1st10,twri1st11}, whereas non-resonant interaction was considered in ref. \cite{twnri1st4}. Similarly, in presence of group velocity dispersion resonant interaction has been investigated in refs. \cite{twri1st,twri1st2,twri1st10} and non-resonant interaction was considered in refs. \cite{twnri2nd,twnri2nd2,huang}.
	
	Three wave resonant and non-resonant interactions can be described by the nonlinear Schr\"odinger equation. It has been solved by the inverse scattering method \cite{ablowitz}, numerical method \cite{numecal.solutions,xie}, variation method \cite{variationa.method} and direct ansatz method \cite{huang,ansatz,xie12}. The stability analysis for the non-degenerate soliton solutions have been investigated in \cite{kref96,xie}. In the year $2011$, Conforti, Baronio and Degasperis \cite{twri1st10} presented that dark-dark-dark (DDD) solutions of TWI are always unstable and checked their stability by modulational instability gain. In this work, we consider a second order $(1+1)$-dimensional TWI with quadratic $\chi^{(2)}$ nonlinearities \cite{a.p.and.p.a16.2,kref96,numecal.solutions,xie,variationa.method,xie12,xie6,xie10}. The nonlinear Schr\"odinger equation has self-consistent periodic solution and it is expressed by Jacobi elliptic functions such as cnoidal, snoidal and dnoidal \cite{carr,malomed,nath.opt,nath.epjp}. The multicomponent self-consistent periodic solution has been defined in nonlinear medium with quadratic nonlinearity which is used in nonlinear optics for parametric frequency conversion \cite{xx}. Stable and unstable modes of three wave resonant interaction in presence of $\mathcal{PT}$-symmetric multi-well Scarf-II potential have been shown in \cite{shen2018}. In this paper we have considered the three-wave interaction with group velocity dispersion and wave number mismatch $\Delta k\ne0$. Different kind of solitons solutions of TWI have been found. The most important is $sech()$ type soliton solution and it was considered for ultrashort pulses \cite{twnri2nd} and optical switching \cite{twnri2nd2}, whereas $a+b\,sech() \tanh() +c\,sech^2()$ type solution was defined in \cite{huang}. The simplest $sech^2()$ type soliton interaction investigated for non-degenerate TWI in refs. \cite{xie,xie12}. Similarly, the dark-dark-dark soliton interaction was defined in ref. \cite{twri1st10}. The first order and second order three wave resonant and non-resonant interaction have constant magnitude solutions.
	
	Our current investigation contributes to the following aspects: we have discovered modulated periodic amplitude solutions expressed in terms of Jacobi elliptic functions, specifically of the form $a+b\,cn^2(\a t,\nu)$ or $a+b \, sn(\a t,\nu)\, cn(\a t,\nu)$, where, $a$ and $b$ are real constants, and $0<\nu\le 1$ represents the elliptic parameter. In the limiting case, $\nu\rightarrow1$ periodic solution approaches to the soliton solution. To achieve this, we conduct analytical and numerical techniques on solutions, considering both cases of $0<\nu< 1$ and $\nu\rightarrow 1$. Our investigation establishes the specific necessary conditions for the existence of self-consistent periodic as well as soliton solutions. From these propose self-consistent periodic solutions one can generate soliton solutions and they were already considered in the literature \cite{twnri2nd,twri1st10,twnri2nd2,huang,xie,xie12}. Next, we investigate the linear stability analysis of periodic and soliton solutions for certain parameter values. As numerical techniques, we use the finite differences, 4th-order Runge-Kutta \cite{numerical}, Crank-Nicholson, Pseudospectral and Fourier collocation methods \cite{jy}. Then we find the stable and unstable modes of periodic solution and one/two hump bright soliton solutions. Next, we investigate the effect of group velocity dispersion and wave number mismatch with respect to the stable and unstable modes. The scope of our investigation pertains to nonlinear optics, as quadratic nonlinear optical media provide unique opportunities for exploring new types of solutions and their stability.
	
	Our paper is organized as follows: in Section \ref{sec2.model}, we show the existence of periodic and soliton solutions of three wave interacting system. In Section \ref{sec3.stability}, we analyse the linear stability of periodic, hyperbolic and constant wave solutions. In Section \ref{results}, we discuss stable and unstable modes of different solutions and compare numerical results corresponding to analytical forms in presence of group velocity dispersion and wave vector mismatch. Finally, in section \ref{conclusion} some concluding remarks are given.	
	%%%%%%%%%%%%%%%%%%%%%%%%%%%%%%%%%%%%%%%%%%%%%%%%%%%%%%%%%%%%%%%%%%%%%%%%%%%%%%
	\section{Theoretical model and solutions}\label{sec2.model}
	We start with the fundamental and second harmonic waves which propagates along $z$ direction and confined in transverse directions \cite{twnri2nd,twnri2nd2,a.p.and.p.a,kref96,kref97,twri1st10,huang} in the form
	\beq \label{Eq.twi}
	\ba{l}
	i\left(\f{\partial }{\partial z}+\f{1}{v_1}\f{\partial}{\partial t}\right)F_1-g_1\f{\partial^2 F_1}{\partial t^2}+\sigma F_2^*F_3e^{i\Delta k z}=0,\\
	i\left(\f{\partial }{\partial z}+\f{1}{v_2}\f{\partial}{\partial t}\right)F_2-g_2\f{\partial^2 F_2}{\partial t^2}+\sigma \f{\om_2}{\om_1}F_3F_1^*e^{i\Delta k z}=0,\\
	i\left(\f{\partial }{\partial z}+\f{1}{v_3}\f{\partial}{\partial t}\right)F_3-g_3\f{\partial^2 F_3}{\partial t^2}+\sigma \f{\om_3}{\om_1}F_1F_2e^{-i\Delta k z}=0,
	\ea
	\eeq
	where, centre frequencies $\om_j$; group velocities $v_j$; second-order dispersion coefficients, i.e. group-velocity dispersion $g_j$; wave numbers $k_j= \f{n_j\om_j}{c}$, $c$ denotes the speed of light, $n_j$ represents the refractive index, nonlinear coupling constants is $\sigma \thickapprox \f{2\pi \xi_{nl} \om_1^2}{k_1 c^2} $, $\xi_{nl}$ represents the nonlinear dielectric susceptibility; and wave vector mismatch is $\Delta k=k_3-k_1-k_2$. The usual TWI equations are found to be completely integrable which occurs in Eq.(\ref{Eq.twi}) when $g_1, g_2$ and $g_3$ are taken zero. We assume the solution of Eq.(\ref{Eq.twi}) in the following form
	\beq\label{F=Aetheta}
	\ba{l}
	F_j=A_j(t)e^{i(\theta_j(t) +\mu_j z)},~~~j=1,2,3 %,~~~p=\Omega\,t'-K z'
	\ea
	\eeq
	where $A_j$'s are real periodic amplitude and $\mu_j$'s are real propagation constants and $\theta_j$ are phase functions of $t$. We substitute Eq.(\ref{F=Aetheta}) into Eq.(\ref{Eq.twi}) and equate the real and imaginary parts, then we obtain
	%\numparts
	\beq \label{Eq.sol.cne0}
	\ba{l}
	g_1\f{\partial^2 A_1}{\partial t^2}+\f{A_1}{2g_1v_1^2}-g_1A_1\left(\f{1}{2g_1v_1}+\f{c_1}{A_1^2}\right)^2+\f{c_1}{v_1g_1}-\sigma_1A_3A_2=-\mu_1 A_1,\\
	g_2\f{\partial^2 A_2}{\partial t^2}+\f{A_2}{2g_2v_2^2}-g_2A_2\left(\f{1}{2g_2v_2}+\f{c_2}{A_2^2}\right)^2+\f{c_2}{v_2g_2}-\sigma_2A_3A_1=-\mu_2 A_2,\\
	g_3\f{\partial^2 A_3}{\partial t^2}+\f{A_3}{2g_3v_3^2}-g_3A_3\left(\f{1}{2g_3v_3}+\f{c_3}{A_3^2}\right)^2+\f{c_3}{v_3g_3}-\sigma_3A_1A_2=-\mu_3 A_3,%\label{Eq.sol.cne0c}
	%\end{eqnarray}
	\ea
	\eeq
	%\endnumparts
	and
	%\numparts
	\beq \label{theta}
	\theta_j(t)=\f{t}{2g_jv_j}+c_j\int_0^t\f{1}{A_j^2(t')}dt',~
	\theta_3=\theta_2+\theta_1,
	\eeq
	\beq\label{mu123}
	\mu_3=\mu_2+\mu_1-\Delta k,
	\eeq
	where $c_j$'s are constant of integration, and we have used the notations
	$\sigma_1=\sigma, \sigma_2=\sigma\f{\om_2}{\om_1},\sigma_3=\sigma\f{\om_3}{\om_1}$, $\sigma_j$'s are real constants. For non-zero $c_j(j=1,2,3)$, the analytical solution of the Eq.(\ref{Eq.sol.cne0}) is complicated. So for the sake of simplicity here we have chosen $c_j=0,~j=1,2,3$. In this case, Eq.(\ref{Eq.sol.cne0}), becomes
	%\numparts
	\beq
	\ba{l}\label{Eq.sol.c0}
	g_1\f{\partial^2 A_1}{\partial t^2}+(\mu_1+\f{1}{4g_1v_1^2})A_1=\sigma_1A_3A_2,\\
	g_2\f{\partial^2 A_2}{\partial t^2}+(\mu_2+\f{1}{4g_2v_2^2})A_2=\sigma_2A_3A_1,\\
	g_3\f{\partial^2 A_3}{\partial t^2}+(\mu_3+\f{1}{4g_3v_3^2})A_3=\sigma_3A_1A_2,
	\ea
	\eeq
	and
	\beq
	\label{g3}
	g_3=\f{1}{v_3\left(\f{1}{g_1v_1}+\f{1}{g_2v_2}\right)}.
	\eeq
	The solution of Eq.(\ref{Eq.sol.c0}) has been solved numerically \cite{numecal.solutions,xie} and approximately by means of variational method \cite{variationa.method}. In the present context, we find the exact solutions of Eq.(\ref{Eq.sol.c0}) in terms of periodic Jacobi elliptic functions and hyperbolic solutions which are either bright or dark solitons and then we utilize the information of $A_1,A_2$, and $A_3$ for the linear stability analysis. To find the exact solutions of Eq.(\ref{Eq.sol.c0}), we will use direct ansatz method \cite{huang,ansatz,xie12} and discuss in the next two subsections \ref{sec.ansatz1} and \ref{sec.ansatz2}.
	%%%%%%%%%%%%%%%%%%%%%%%%%%%%%%%%%%%%%%%%%%%%%%%%%%%%%%%%%%%%%%%%%%%%%%%%%%%%%%
	\subsection{Ansatz I: Self consistent periodic solution}\label{sec.ansatz1}
\begin{figure}[h] 
		\centering%	1
		\includegraphics[width=18cm,height=12cm]{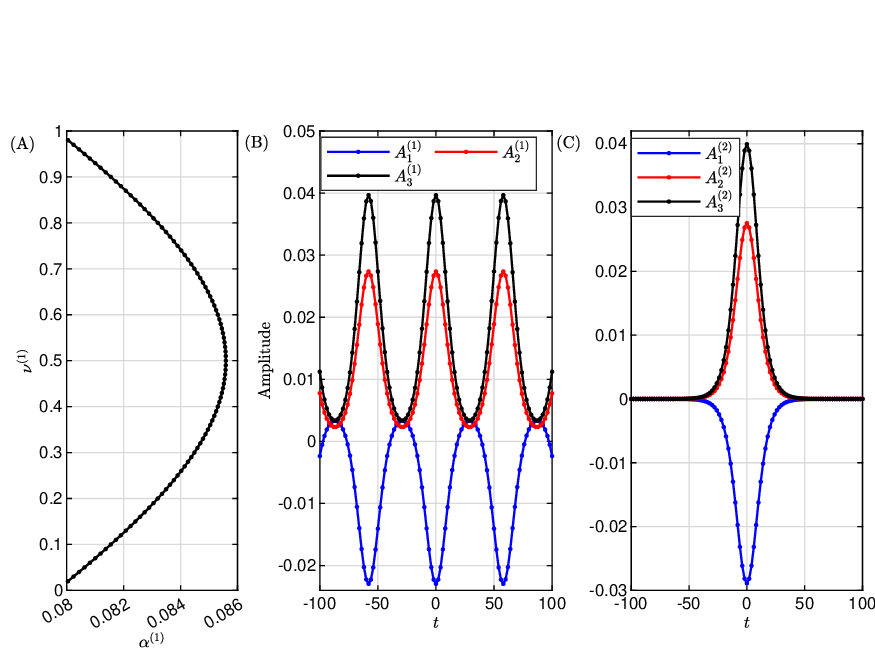}
		\caption{\label{fig1}(A) Plot of the relation between $\alpha^{(1)}$ and $\nu^{(1)}$ from Eq.(\ref{An1.alpha-nu}) for the parameters $\Delta k=-0.04, g_1=1, g_2=1.1$. For $\omega_1=\omega_2=\omega_3=1, \sigma=1, v_1=v_2=v_3=1$ and the corresponding periodic modulated amplitudes are plotted in (B) of Eq.(\ref{ansatz1}) for $\nu^{(1)}=0.85$ and the soliton solution plot in (C) of Eq.(\ref{ansatz1.nu1}).}
	\end{figure}

	Let us take the ansatz
	\beq\label{ansatz1}
	A_j^{(1)}(t)=a_j^{(1)}+b_j^{(1)} cn^2(\alpha^{(1)} t,\nu^{(1)}),~~~j=1,2,3
	\eeq
	where $cn(\alpha^{(1)} t,\nu^{(1)})$ is the Jacobi elliptic cnoidal function with period $\f{4K(\nu^{(1)})}{\alpha^{(1)}}$, and $a_j^{(1)},b_j^{(1)}$ are real constants yet to be determined with \cite{stegun}
	\beq
	K(\nu^{(1)})=\int_0^{\f{\pi}{2}}\f{1}{\sqrt{1-\nu^{(1)}\sin^2\theta}}\,d\theta, %eft(-1,1\right)
	\eeq
	real quarter-period $K(\nu^{(1)})$ of the Jacobi elliptic functions $sn$, $cn$ and $0<\nu^{(1)}<1$. Therefore, the amplitudes $A_1^{(1)},A_2^{(1)},A_3^{(1)}$ are periodic functions of $t$ with period $\f{2K(\nu^{(1)})}{\alpha^{(1)}}$. % If $\alpha=\f{K(\nu)}{\pi}$, then $A_j(j=1,2,3)$'s are periodic functions in $t$ with period $2\pi$.
	Now $e^{i\theta_j(t)}$ is periodic function with period $4g_jv_j\pi$ for $j=1,2,3$. Hence the solutions $F_j$'s given by Eq.(\ref{F=Aetheta}) are periodic functions of $t$. Now, substituting Eq.(\ref{ansatz1}) in Eq.(\ref{Eq.sol.c0}), we obtain
	\beq\label{bj}
	\ba{l}
	\left[b_1^{(1)}\right]^2=\f{36\left[\nu^{(1)}\right]^2\left[\alpha^{(1)}\right]^4g_2g_3}{\sigma_2\sigma_3},~ \left[b_2^{(1)}\right]^2=\f{36\left[\nu^{(1)}\right]^2\left[\alpha^{(1)}\right]^4g_1g_3}{\sigma_1\sigma_3},~ \left[b_3^{(1)}\right]^2=\f{36\left[\nu^{(1)}\right]^2\left[\alpha^{(1)}\right]^4g_1g_2}{\sigma_1\sigma_2}.
	\ea
	\eeq
	It is to be noted that, one of the $b^{(1)}_1,b^{(1)}_2,b^{(1)}_3$'s is negative, together with either $g_1,g_2,g_3>0$ or $g_1,g_2,g_3<0$. Therefore, without loss of generality, we can choose $b^{(1)}_1<0$, then we obtain
	%\numparts
	\beq\label{b1nb2b3p}
	b_1^{(1)}=-6\nu^{(1)}\left[\alpha^{(1)}\right]^2\sqrt{\f{g_2g_3}{\sigma_2\sigma_3}},~ b_2^{(1)}=6\nu^{(1)}\left[\alpha^{(1)}\right]^2\sqrt{\f{g_3g_1}{\sigma_3\sigma_1}},~ b_3^{(1)}=6\nu^{(1)}\left[\alpha^{(1)}\right]^2\sqrt{\f{g_1g_2}{\sigma_1\sigma_2}},
	\eeq
	and 
	\beq\label{aj}
	\ba{l}
	\ds a_1^{(1)}=\f{18\left[\nu^{(1)}\right]^2\left[\alpha^{(1)}\right]^4g_1g_2g_3}{\sigma_1\sigma_2\sigma_3b^{(1)}_2b^{(1)}_3}\left(-\f{\mu^{(1)}_1+\xi_1}{g_1}+\f{\mu^{(1)}_2+\xi_2}{g_2}+\f{\mu^{(1)}_3+\xi_3}{g_3}\right),\\
	\ds a_2^{(1)}=\f{18\left[\nu^{(1)}\right]^2\left[\alpha^{(1)}\right]^4g_1g_2g_3}{\sigma_1\sigma_2\sigma_3b^{(1)}_3b^{(1)}_1}\left(\f{\mu^{(1)}_1+\xi_1}{g_1}-\f{\mu^{(1)}_2+\xi_2}{g_2}+\f{\mu^{(1)}_3+\xi_3}{g_3}\right),\\
	\ds a_3^{(1)}=\f{18\left[\nu^{(1)}\right]^2\left[\alpha^{(1)}\right]^4g_1g_2g_3}{\sigma_1\sigma_2\sigma_3b^{(1)}_1b^{(1)}_2}\left(\f{\mu^{(1)}_1+\xi_1}{g_1}+\f{\mu^{(1)}_2+\xi_2}{g_2}-\f{\mu^{(1)}_3+\xi_3}{g_3}\right),
	\ea
	\eeq
	where
	\beq
	\xi_j=\f{1}{4g_jv_j^2}-4g_j\left[\alpha^{(1)}\right]^2\left(1-2\nu^{(1)}\right)\label{xi}.
	\eeq
	In order to find the values of the remaining five parameters $\m^{(1)}_1,\mu^{(1)}_2,\mu^{(1)}_3, \alpha^{(1)},\nu^{(1)}$, we have three relations %and $b_j$'s and $a_j$'s are satisfying the relations
	\beq\label{muFirst}
	\ba{l}
	2\left(\mu^{(1)}_1+\f{1}{4g_1v_1^2}\right)\left\{-\f{\mu^{(1)}_1}{g_1}+\f{\mu^{(1)}_1}{g_3}+\f{\mu^{(1)}_2}{g_2}+\f{\mu^{(1)}_2}{g_3}-\f{\Delta k}{g_3}+\f{1}{4}\left(-\f{1}{g_1^2v_1^2}+\f{1}{g_2^2v_2^2}+\f{1}{g_3^2v_3^2}\right)-4\left[\alpha^{(1)}\right]^2(1-2\nu^{(1)})\right\}\\
	~~=g_1\left\{\f{\mu^{(1)}_1}{g_1}+\f{\mu^{(1)}_1}{g_3}-\f{\mu^{(1)}_2}{g_2}+\f{\mu^{(1)}_2}{g_3}-\f{\Delta k}{g_3}+\f{1}{4}\left(\f{1}{g_1^2v_1^2}-\f{1}{g_2^2v_2^2}+\f{1}{g_3^2v_3^2}\right)-4\left[\alpha^{(1)}\right]^2(1-2\nu^{(1)})\right\}\\
	~~\times\left\{\f{\mu^{(1)}_1}{g_1}-\f{\mu^{(1)}_1}{g_3}+\f{\mu^{(1)}_2}{g_2}-\f{\mu^{(1)}_2}{g_3}+\f{\Delta k}{g_3}+\f{1}{4}\left(\f{1}{g_1^2v_1^2}+\f{1}{g_2^2v_2^2}-\f{1}{g_3^2v_3^2}\right)-4\left[\alpha^{(1)}\right]^2(1-2\nu^{(1)})\right\}
	+48g_1\left[\alpha^{(1)}\right]^4\nu^{(1)}(1-\nu^{(1)}),\\
	2\left(\mu^{(1)}_2+\f{1}{4g_2v_2^2}\right)\left\{\f{\mu^{(1)}_1}{g_1}+\f{\mu^{(1)}_1}{g_3}-\f{\mu^{(1)}_2}{g_2}+\f{\mu^{(1)}_2}{g_3}-\f{\Delta k}{g_3}+\f{1}{4}\left(\f{1}{g_1^2v_1^2}-\f{1}{g_2^2v_2^2}+\f{1}{g_3^2v_3^2}\right)-4\left[\alpha^{(1)}\right]^2(1-2\nu^{(1)})\right\}\nonumber\\
	~~=g_2\left\{\f{\mu^{(1)}_1}{g_1}-\f{\mu^{(1)}_1}{g_3}+\f{\mu^{(1)}_2}{g_2}-\f{\mu^{(1)}_2}{g_3}+\f{\Delta k}{g_3}+\f{1}{4}\left(\f{1}{g_1^2v_1^2}+\f{1}{g_2^2v_2^2}-\f{1}{g_3^2v_3^2}\right)-4\left[\alpha^{(1)}\right]^2(1-2\nu^{(1)})\right\}\nonumber\\
	~\times \left\{-\f{\mu^{(1)}_1}{g_1}+\f{\mu^{(1)}_1}{g_3}+\f{\mu^{(1)}_2}{g_2}+\f{\mu^{(1)}_2}{g_3}-\f{\Delta k}{g_3}+\f{1}{4}\left(-\f{1}{g_1^2v_1^2}+\f{1}{g_2^2v_2^2}+\f{1}{g_3^2v_3^2}\right)-4\left[\alpha^{(1)}\right]^2(1-2\nu^{(1)})\right\}
	+48g_2\left[\alpha^{(1)}\right]^4\nu^{(1)}(1-\nu^{(1)}),\\
	2\left(\mu^{(1)}_1+\mu^{(1)}_2-\Delta k+\f{1}{4g_3v_3^2}\right)\left\{\f{\mu^{(1)}_1}{g_1}-\f{\mu^{(1)}_1}{g_3}+\f{\mu^{(1)}_2}{g_2}-\f{\mu^{(1)}_2}{g_3}+\f{\Delta k}{g_3}+\f{1}{4}\left(\f{1}{g_1^2v_1^2}+\f{1}{g_2^2v_2^2}-\f{1}{g_3^2v_3^2}\right)\right.-4\left[\alpha^{(1)}\right]^2(1-2\nu^{(1)})\Big\}\\
	~~=g_3\left\{-\f{\mu^{(1)}_1}{g_1}+\f{\mu^{(1)}_1}{g_3}+\f{\mu^{(1)}_2}{g_2}+\f{\mu^{(1)}_2}{g_3}-\f{\Delta k}{g_3}+\f{1}{4}\left(-\f{1}{g_1^2v_1^2}+\f{1}{g_2^2v_2^2}+\f{1}{g_3^2v_3^2}\right)-4\left[\alpha^{(1)}\right]^2(1-2\nu^{(1)})\right\}\\
	~~\times\left\{\f{\mu^{(1)}_1}{g_1}+\f{\mu^{(1)}_1}{g_3}-\f{\mu^{(1)}_2}{g_2}+\f{\mu^{(1)}_2}{g_3}-\f{\Delta k}{g_3}+\f{1}{4}\left(\f{1}{g_1^2v_1^2}-\f{1}{g_2^2v_2^2}+\f{1}{g_3^2v_3^2}\right)-4\left[\alpha^{(1)}\right]^2(1-2\nu^{(1)})\right\}
	+48g_3\left[\alpha^{(1)}\right]^4\nu^{(1)}(1-\nu^{(1)}),
	\ea
	\eeq
	and satisfying equation (\ref{mu123}). Therefore, the parameters $\m^{(1)}_1,\mu^{(1)}_2,\mu^{(1)}_3, \alpha^{(1)},\nu^{(1)}$ and $\Delta k$ are functionally connected. One can find a relation among the parameters. In this paper, we show the condition for existence of the solution of these parameters numerically for some particular set of values. We can solve the first and second equations of (\ref{muFirst}) for $\mu^{(1)}_1, \mu^{(1)}_2$ in terms of $\alpha^{(1)}, \nu^{(1)}, g_1,g_2$ for different values of $v_1,v_2,v_3,\sigma,\omega_1,\omega_2,\omega_3$ and $\Delta k$. In particular, for $v_1=v_2=v_3=1, \omega_1=\omega_2=\omega_3=1, \sigma=1$, we obtain
	\beq
	\ba{ll}
	\mu^{(1)}_1&=\frac{-16\sqrt{[\alpha^{(1)}]^4g_1^8g_2^2\left(g_1+g_2\right)^4\left(1-\nu^{(1)}+[\nu^{(1)}]^2\right)}+4\Delta kg_1^6+g_1^5\left(12\Delta kg_2-1\right)+4g_2g_1^4 \left(3\Delta kg_2-1\right)+2g_2^2g_1^3\left(2\Delta kg_2-3\right)-4g_2^3g_1^2-g_2^4g_1}{4g_1^2\left(g_1+g_2\right)^4},\\
	\mu^{(1)}_2&=\frac{-16g_2^2\sqrt{[\alpha^{(1)}]^4g_1^8g_2^2\left(g_1+g_2\right)^4\left(1-\nu^{(1)}+[\nu^{(1)}]^2\right)}-g_1^7+4g_2g_1^6\left(\Delta kg_2-1\right)+6g_2^2g_1^5\left(2 \Delta kg_2-1\right)+4g_2^3g_1^4\left(3\Delta kg_2-1\right)+g_2^4g_1^3\left(4\Delta kg_2-1\right)}{4g_1^3g_2\left(g_1+g_2\right)^4}
	\ea
	\eeq
	and substitute these values of $\mu^{(1)}_1,\mu^{(1)}_2$ into the last equation of (\ref{muFirst}) and obtain a relation between $\alpha^{(1)}, \nu^{(1)}$ as
	\beq\label{An1.alpha-nu}
	\ba{r}
	\Delta k\sqrt{[\alpha^{(1)}]^4g_1^8g_2^2\left(g_1+g_2\right)^4\left(1-\nu^{(1)}+[\nu^{(1)}]^2\right)}+4[\alpha^{(1)}]^4g_2g_1^7\left(1-\nu^{(1)}+[\nu^{(1)}]^2\right)+8[\alpha^{(1)}]^4g_2^2 g_1^6\left(1-\nu^{(1)}+[\nu^{(1)}]^2\right)\\+8[\alpha^{(1)}]^4g_2^3g_1^5\left(1-\nu^{(1)}+[\nu^{(1)}]^2\right)
	+4[\alpha^{(1)}]^4g_2^4g_1^4\left(1-\nu^{(1)}+[\nu^{(1)}]^2\right)=0.
	\ea
	\eeq
	Therefore, the first set of periodic solutions defined in (\ref{ansatz1}) exists. In Fig. \ref{fig1} (A), we have plotted the existence of solution (\ref{ansatz1}) and the corresponding periodic solution is plotted in Fig. \ref{fig1} (B).
	\subsection{One hump bright soliton solution}
	In the limiting case $\nu^{(1)}\rightarrow1$, the periodic solutions (\ref{ansatz1}) become hyperbolic and the corresponding second set of hyperbolic solutions exist and defined by
	%\numparts
	\beq\label{ansatz1.nu1}
	\ba{l}
	A_1^{(2)}(t)=a_1^{(2)}-6\left[\alpha^{(2)}\right]^2\sqrt{\f{g_2g_3}{\sigma_2\sigma_3}}\,sech^2\alpha^{(2)} t\label{b1.nu1},\\ A_2^{(2)}(t)=a_2^{(2)}+6\left[\alpha^{(2)}\right]^2\sqrt{\f{g_3g_1}{\sigma_3\sigma_1}}\,sech^2\alpha^{(2)} t\label{b2.nu1},\\ A_3^{(2)}(t)=a_3^{(2)}+6\left[\alpha^{(2)}\right]^2\sqrt{\f{g_1g_2}{\sigma_1\sigma_2}}\,sech^2\alpha^{(2)} t,
	\ea
	\eeq
	where $a_j^{(2)}, j=1,2,3$ and other parameters are obtained from Eqs. (\ref{aj})-(\ref{An1.alpha-nu}) for $\nu^{(1)}=1$.\\
	The hyperbolic solutions of the second set are plotted in Fig. \ref{fig1} (C). From Fig. \ref{fig1} (C), one can observe that the components of the hyperbolic solutions are bright solitons.
	
	\subsection{Ansatz II: Self consistent periodic solution}\label{sec.ansatz2}
	Next we have considered the ansatz of the form
	\beq\label{ansatz2A1}
	\ba{ll}
	A_1^{(3)}(t)=a_1^{(3)}+b_1^{(3)}\,cn^2(\alpha^{(3)} t,\nu^{(3)}),\\
	A_j^{(3)}(t)=a_j^{(3)}+b_j^{(3)}\,sn(\alpha^{(3)} t,\nu^{(3)})\,cn(\alpha^{(3)} t,\nu^{(3)}),~~j=2,3
	\ea
	\eeq
	where $sn(\alpha^{(3)} t,\nu^{(3)})$ is the Jacobi elliptic snoidal function, $0<\nu^{(3)}<1$ and $a^{(3)}_j, b^{(3)}_j$ are real constants left to be determined. Therefore $A^{(3)}_1$ is a periodic function in $t$ with period $\f{2K(\nu^{(3)})}{\alpha^{(3)}}$ and $A^{(3)}_2,~A^{(3)}_3$ are periodic functions in $t$ with the same period $\f{4K(\nu^{(3)})}{\alpha^{(3)}}$. Hence the solutions $F_1,F_2,F_3$ to be periodic function in $t$. Substituting Eq.(\ref{ansatz2A1}) into Eqs.(\ref{Eq.sol.c0}) we obtain
	\beq\label{bj.2g1.less0}
	\left.\ba{l}b_1^{(3)}=-6\nu^{(3)}\left[\alpha^{(3)}\right]^2\sqrt{\f{g_2g_3}{\sigma_2\sigma_3}},~b_2^{(3)}=6\nu^{(3)}\left[\alpha^{(3)}\right]^2\sqrt{-\f{g_1g_3}{\sigma_1\sigma_3}},~b_3^{(3)}=6\nu^{(3)}\left[\alpha^{(3)}\right]^2\sqrt{-\f{g_1g_2}{\sigma_1\sigma_2}},\\
	g_1<0,g_2,g_3>0\ea\right\},
	\eeq
	or
	\beq\label{bj.2g1.greater0}
	\left.\ba{l}b_1^{(3)}=6\nu^{(3)}\left[\alpha^{(3)}\right]^2\sqrt{\f{g_2g_3}{\sigma_2\sigma_3}},~b_2^{(3)}=6\nu^{(3)}\left[\alpha^{(3)}\right]^2\sqrt{-\f{g_1g_3}{\sigma_1\sigma_3}},~b_3^{(3)}=6\nu^{(3)}\left[\alpha^{(3)}\right]^2\sqrt{-\f{g_1g_2}{\sigma_1\sigma_2}},\\
	g_1>0,g_2,g_3<0\ea\right\},
	\eeq
	and
	\beq\label{a1.2}
	a_1^{(3)}=-\f{(1-\nu^{(3)})b_1^{(3)}}{2-\nu^{(3)}},~a_2^{(3)}=a_3^{(3)}=0.
	\eeq
	The propagation constants $\mu_j$'s are defined by
	\beq\label{mu.sol3}
	\ba{l}
	\ds\mu^{(3)}_1=g_1\left[\alpha^{(3)}\right]^2(4-2\nu^{(3)})-\ds\f{1}{4g_1v_1^2},\\
	\ds\mu^{(3)}_2=\ds g_2\left[\alpha^{(3)}\right]^2 \f{8-8\nu^{(3)}-\left[\nu^{(3)}\right]^2}{2-\nu^{(3)}}-\ds\f{1}{4g_2v_2^2},\\
	\ds\mu^{(3)}_3=g_3\left[\alpha^{(3)}\right]^2\f{8-8\nu^{(3)}-\left[\nu^{(3)}\right]^2}{2-\nu^{(3)}}-\ds\f{1}{4g_3v_3^2},
	\ea
	\eeq
	and the relation between remaining two parameters $\alpha^{(3)}$, $\nu^{(3)}$ are obtained from Eq.(\ref{mu123}) and it is defined by 
	\beq\label{alphanu.2}
	\left[\alpha^{(3)}\right]^2\left\{(g_3-g_2)\f{8-8\nu^{(3)}-\left[\nu^{(3)}\right]^2}{2-\nu^{(3)}}-g_1\left(4-2\nu^{(3)}\right)\right\}=\f{1}{4}\left(\f{1}{g_3v_3^2}-\f{1}{g_2v_2^2}-\f{1}{g_1v_1^2}\right)-\Delta k.
	\eeq
	Therefore, the third set of periodic solutions of the form (\ref{ansatz2A1}) exist, if the parameters satisfy relation (\ref{alphanu.2}). In particular, for $\Delta k=0.008, v_1=-2\times10^8, v_2=10^8, v_3=10^8,g_1=-0.001, g_2=1.3$, the relation (\ref{alphanu.2}) is valid for $0<\nu^{(3)}<0.9$ and the corresponding valid region is shown in figure \ref{fig2}(A) and the amplitudes of $A_j^{(3)}(j=1,2,3)$ for $\nu^{(3)}=0.01$ are plotted in figure \ref{fig2}(B) .
	\subsection{One and two hump bright soliton solution}
\begin{figure}[ht] % 2 Fig for relation between $\alpha$ and $\nu$ and 2D plot of solution A^{(3)} and A^{(4)}
		\centering	
		\includegraphics[width=18cm,height=12cm]{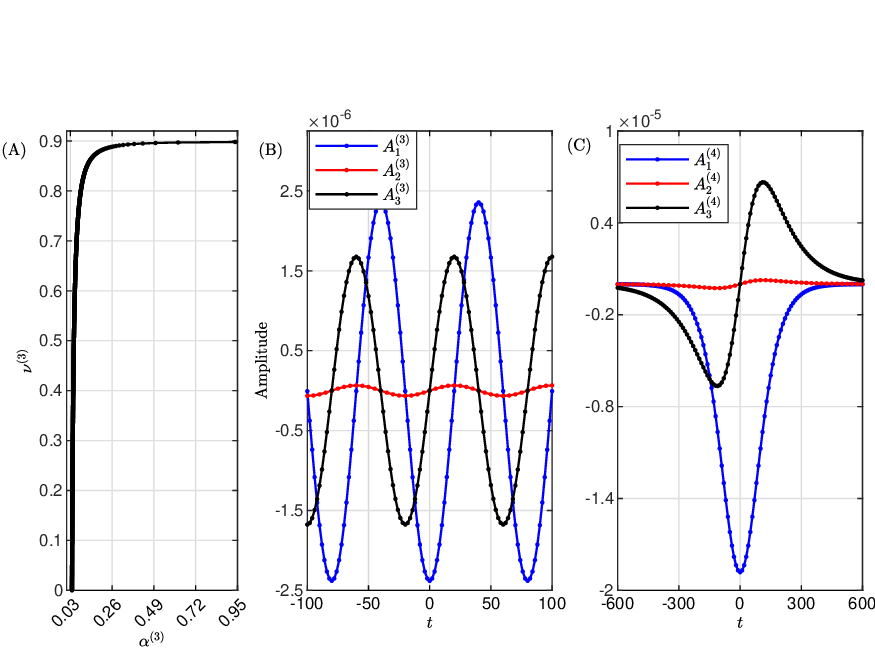}
		\caption{\label{fig2}(A) Plot of the relation between $\alpha^{(3)}$ and $\nu^{(3)}$ from Eq.(\ref{alphanu.2}) for the parameters $\Delta k=0.008, g_1=-0.001, g_2=1.3$. For $v_1=-2\times10^8, v_2=v_3=10^8, \omega_1=\omega_2=\omega_3=1, \sigma=1$ and the corresponding modulated periodic amplitudes are plotted in (B) of Eq.(\ref{ansatz2A1}) for $\nu^{(3)}=0.01, \Delta k=0.008$ and the soliton solutions are plotted in (C) of Eq.(\ref{Ansatz2.A1.nu1}) for $\Delta k=-0.00008$.}
	\end{figure}

	In the limiting case, when $\nu^{(3)}\rightarrow1$, periodic solutions given by Eq.(\ref{ansatz2A1}) become
	\beq\label{Ansatz2.A1.nu1}
	\ba{ll}
	A_1^{(4)}(t)=b_1^{(4)}\,sech^2\alpha^{(4)} t,\\
	A_j^{(4)}(t)=b_j^{(4)}\,sech\alpha^{(4)} t\tanh\alpha^{(4)} t,~~j=2,3,
	\ea
	\eeq
	where $b_j^{(4)}, j=1,2,3$ and other parameters are obtained from Eqs. (\ref{bj.2g1.greater0}), (\ref{mu.sol3}) and (\ref{alphanu.2}) for $\nu^{(3)}=1$.\\
	In Fig. \ref{fig2}(C), we have plotted the hyperbolic solutions (\ref{Ansatz2.A1.nu1}) of the fourth set, for $\Delta k=-0.00008, v_1=-2\times10^8, v_2=10^8, v_3=10^8,g_1=-0.001, g_2=1.3$. From Fig. \ref{fig2}(C), one can observe that the components of hyperbolic solution are one hump and two humps bright solitons.
	
	\subsection{Constant magnitude wave solution}
	In a special case, the constant magnitude i.e., fifth set of plane wave solutions
	\beq\label{a5}
	F_j^{(5)}=a_j^{(5)}e^{i(\theta_j(t) +\mu_j^{(5)} z)},~~~j=1,2,3
	\eeq
	exist and the values of $a_j^{(5)}$'s are defined by
	%\numparts
	\beq\label{planewave1}
	\ba{l}
	\left[a_1^{(5)}\right]^2=\f{1}{\sigma_2\sigma_3}\left(\mu^{(5)}_2+\f{1}{4g_2v_2^2}\right)\left(\mu_3^{(5)}+\f{1}{4g_3v_3^2}\right),\\
	\left[a_2^{(5)}\right]^2=\f{1}{\sigma_3\sigma_1}\left(\mu_3^{(5)}+\f{1}{4g_3v_3^2}\right)\left(\mu^{(5)}_1+\f{1}{4g_1v_1^2}\right),\\
	\left[a_3^{(5)}\right]^2=\f{1}{\sigma_1\sigma_2}\left(\mu_1^{(5)}+\f{1}{4g_1v_1^2}\right)\left(\mu^{(5)}_2+\f{1}{4g_2v_2^2}\right),
	\ea
	\eeq
	where $\mu_3^{(5)}=\mu_2^{(5)}+\mu_1^{(5)}-\Delta k $ and each term of the right hand side of Eq.(\ref{planewave1}) are positive. 
	
	In the limiting case, $\nu^{(1)}\rightarrow0$, the trigonometric solutions of ansatz I in the form $A_j^{(6)}(t)=a_j^{(6)}+b_j^{(6)}\cos^2\alpha t,~~~j=1,2,3$ do not exist for non-zero $g_j$'s and $b_j$'s. Similarly, in the limiting case, $\nu^{(3)}\rightarrow0$, the trigonometric solutions of ansatz II in the form $A_1^{(7)}(t,\a,0)=a_1^{(7)}+b_1^{(7)}\cos^2\alpha t$, $A_j^{(7)}(t)=a_j^{(7)}+b_j^{(7)}\sin\alpha t\cos\alpha t,j=2,3$ do not exist.
	
	%%%%%%%%%%%%%%%%%%%%%%%%%%%%%%%%%%%%%%%%%%%%%%%%%%%%%%
	\section{Linear stability analysis of multiple solutions}\label{sec3.stability}
	The stability properties of first four sets of solutions are tested in direct numerical simulations as well as within the framework of linear stability analysis. For the linear stability analysis we have considered small perturbation to the stationary solution of the form
	\beq\label{perturbation}
	F_j^{(i)}(t,z)=\left(A_j^{(i)}(t)+\epsilon U_j(t,z)+O(\epsilon^2)\right)e^{i(\theta_j(t)+\mu^{(i)}_j z)},~~~j=1,2,3,~i=1,2,3,4,
	\eeq
	where $\epsilon, |U_1|, |U_2|, |U_3|\ll 1$ are infinitesimal complex perturbations. We substitute the perturbed solution defined in (\ref{perturbation}) into Eq.(\ref{Eq.twi}) and yield the following equations
	\beq\label{EvolutionUz}
	\ba{l}
	i\f{\partial U_1}{\partial z}=L_1^{(i)}U_1+i S_1^{(i)}U_1-\sigma_1\left(A_3^{(i)}U_2^*+A_2^{(i)}U_3\right),\\
	i\f{\partial U_2}{\partial z}=L_2^{(i)}U_2+i S_2^{(i)}U_2-\sigma_2\left(A_3^{(i)}U_1^*+A_1^{(i)}U_3\right),\\
	i\f{\partial U_3}{\partial z}=L_3^{(i)}U_3+i S_3^{(i)}U_3-\sigma_3\left(A_1^{(i)}U_2+A_2^{(i)}U_1\right),
	\ea
	\eeq
	where $^*$ represents the complex conjugate and the linear operators $L_j^{(i)}$ and $S_j^{(i)}$ are defined by
	%\numparts
	\beq\label{operatorLj}
	\left.\ba{l}
	\ds L_j^{(i)}=g_j\f{\partial^2}{\partial t^2}+\f{c_j}{v_j\left[A_j^{(i)}\right]^2}-g_j\left(\f{1}{2g_jv_j}+\f{c_j}{\left[A_j^{(i)}\right]^2}\right)^2+\mu^{(i)}_j+\f{1}{2g_jv_j^2},\\
	S_j^{(i)}=\f{2c_jg_j}{A_j^{(i)}}\f{\partial}{\partial t}\left(\f{1}{A_j^{(i)}}\right)
	\ea\right\},~j=1,2,3.
	\eeq
	%\endnumparts
	To solve Eq.(\ref{EvolutionUz}), we introduce a transformation on $U_j(t,z)$ which is normal mode of small perturbed eigen functions $u_j(t)$ and $v_j(t)$
	\beq\label{transformationU}
	U_j(t,z)=\left(u_j(t)+i\,v_j(t)\right)e^{\lam z},~~~j=1,2,3.
	\eeq
	The eigen function may increase with the perturbation growth rate $\lam$ during propagation. Now we substitute the above transformation into Eq.(\ref{EvolutionUz}) and equate the real and imaginary parts and obtain a linear eigenvalue problem
	\beq\label{Eq.EV.cne0}
	\lambda \left(
	\ba{c}u\\v\ea
	\right)=\left(
	\ba{lc}S^{(i)} & \left[L^{(i)}\right]^+\\\left[L^{(i)}\right]^-& S^{(i)}\ea
	\right)\left(
	\ba{c}u\\v\ea
	\right),
	\eeq
	where $u=\left(u_1~u_2~u_3\right)^T, v=\left(v_1~v_2~v_3\right)^T$, $T$ denotes the transpose and the operators $\left[L^{(i)}\right]^+, \left[L^{(i)}\right]^-, S^{(i)}$ are defined by
	\beq
	\ba{l}
	\left[L^{(i)}\right]^+=L^{(i)}+M^{(i)}_1,~\left[L^{(i)}\right]^-=-L^{(i)}+M^{(i)}_2,\\
	S^{(i)}=(s^{(i)}_{jk})_{3\times3},~
	L^{(i)}=(l^{(i)}_{jk})_{3\times3},
	\ea
	\eeq
	where
	\beq
	\ba{l}
	M_1^{(i)}=\left(\ba{ccc} 0 & \sigma_1A^{(i)}_3 & -\sigma_1A^{(i)}_2\\
	\sigma_2A^{(i)}_3 & 0 & -\sigma_2A^{(i)}_1\\
	-\sigma_3A^{(i)}_2 & -\sigma_3A^{(i)}_1 & 0\ea\right),~
	M_2^{(i)}=M_1^{(i)}+2\left(\ba{ccc} 0 & 0 & \sigma_1A^{(i)}_2\\
	0 & 0 & \sigma_2A^{(i)}_1\\
	\sigma_3A^{(i)}_2 & \sigma_3A^{(i)}_1 & 0\ea\right),
	\ea
	\eeq
	and
	\beq
	\ba{l}
	l^{(i)}_{jk}=\left\{\ba{l}L^{(i)}_j,~~~j=k\\
	0,~~~j\ne k\ea \right\},~s^{(i)}_{jk}=\left\{\ba{l}S^{(i)}_j,~~~j=k\\
	0,~~~j\ne k\ea \right\}.
	\ea
	\eeq
	The eigenvalue problem (\ref{Eq.EV.cne0}) opens a crucial path to analyses linear stability for both periodic and hyperbolic solutions. If the eigenvalue $\lambda$ has a positive real part, then perturbed solution mode is completely unstable as the solution will grow exponentially with $z$. On the other hand, the solution mode becomes completely stable only when real parts of $\lambda$ are zero or negative. It is to be mentioned that, the study of linear stability for non-trivial phase solution $(c_j\ne 0,~j=1,2,3)$ is an open problem, even in solutions of Eq.(\ref{Eq.sol.cne0})  \cite{nath.opt,nath.epjp,carr,malomed}. So, we restrict ourselves $c_j=0$, for $j=1,2,3$. However, all waves of TWI system are described by $c_j=0$ \cite{kref96,kref97}. For $c_j=0$, we obtain $S^{(i)}_j=0$ and the corresponding spectral problem Eq.(\ref{Eq.EV.cne0}) reduces to
	\beq\label{Eq.EV.c0}
	\left(\ba{rrrrrr}
	0&0&0&L_1^{(i)}&\sigma_1A_3^{(i)}&-\sigma_1A_2^{(i)}\\
	0&0&0&\sigma_2A_3^{(i)}&L_2^{(i)}&-\sigma_2A_1^{(i)}\\
	0&0&0&-\sigma_3A_2^{(i)}&-\sigma_3A_1^{(i)}&L_3^{(i)}\\
	-L_1^{(i)}&\sigma_1A_3^{(i)}&\sigma_1A_2^{(i)}u_3&0&0&0\\
	\sigma_2A_3^{(i)}&-L_2^{(i)}&\sigma_2A_1^{(i)}u_3&0&0&0\\
	\sigma_3A_2^{(i)}&\sigma_3A_1^{(i)}&-L_3^{(i)}u_3&0&0&0
	\ea\right)\left(\ba{l}u_1\\u_2\\u_3\\v_1\\v_2\\v_3\ea\right)=\lambda\left(\ba{l}u_1\\u_2\\u_3\\v_1\\v_2\\v_3\ea\right)
	\eeq
	and the linear operators $L_j^{(i)}$'s are defined by
	\beq\label{operatorLj.c0}
	\ba{l}
	L_j^{(i)}=g_j\f{\partial^2}{\partial t^2}+\mu_j^{(i)}+\f{1}{4g_jv_j^2},~~~j=1,2,3,~i=1,2,3,4.
	\ea
	\eeq
	Therefore, if the spectral problem Eq.(\ref{Eq.EV.c0}) has any positive real part of the eigenvalues, then the solution is linearly unstable and the corresponding wave amplitude change its shape as $z$ increases. In case if all eigenvalues of Eq.(\ref{Eq.EV.c0}) are imaginary or zero, then the solution is linearly stable.
	
	To pursue eigenvalues of Eq. (\ref{Eq.EV.c0}) we have utilized Fourier collocation method (FCM). Three waves defined in Eqs. (\ref{ansatz1}) and (\ref{ansatz2A1}) are periodic functions in $t$ with period $T^{(1)}_{j}=\f{2K(\nu^{(1)})}{\alpha^{(1)}} (j=1,2,3)$ and $T^{(3)}_{1}=\f{2K(\nu^{(3)})}{\alpha^{(3)}}$, $T^{(3)}_{j}=\f{4K(\nu^{(3)})}{\alpha^{(3)}} (j=2,3)$.
	Therefore, we express the perturbation vectors $u_j(t)$ and $v_j(t)$ in the spatial domain $\left[0,T^{(i)}\right]$ in the form
	\beq\label{Fourier.uj}
	\ba{l}
	u_j=\sum\limits_{n=-\infty}^{\infty}u_{j,n}e^{ink^{(i)}_0t},~
	v_j=\sum\limits_{n=-\infty}^{\infty}v_{j,n}e^{ink^{(i)}_0t},~j=1,2,3,~i=1,2,3,4,
	\ea
	\eeq
	where $k^{(i)}_0=\f{2\pi}{T^{(i)}}$, and $u_{j,n}$, $v_{j,n}$ are the $n$th Fourier coefficient of $u_j$ and $v_j$ respectively. Let the Fourier series of the solutions $A^{(i)}_j$'s are
	\beq\label{Fourier.Aj}
	A_j^{(i)}=\ds\sum_{m=-\infty}^{\infty}a_{j,m}^{(i)}e^{imk^{(i)}_0t},~~~j=1,2,3,~i=1,2,3,4,
	\eeq
	where $a_{j,m}^{(i)}$ is the $m$th Fourier coefficient of $A^{(i)}_j$. The analytical form of the Fourier series of $A^{(i)}_j (i=1,3; j=1,2,3)$ corresponding to the solutions given by Eq.(\ref{ansatz1}) and Eq.(\ref{ansatz2A1}) can be obtained from the relation \cite{elliptic}
	\beq\label{Fourier.sn2}
	sn^2(x,\nu^{(i)})=\f{1}{\nu^{(i)}}-\f{E(\nu^{(i)})}{\nu^{(i)} K(\nu^{(i)})}-\f{2\pi^2}{\nu^{(i)} K^2(\nu^{(i)})}\sum\limits_{m=1}^{\infty}\f{mq^m}{1-q^{2m}}\cos\left(\f{m\pi x}{K(\nu^{(i)})}\right),~i=1,3,
	\eeq
	where
	\beq
	\ba{l}
	q=e^{-\f{\pi\,K(1-\nu^{(i)})}{K(\nu^{(i)})}},~\ds E(\nu^{(i)})=\int_0^{\f{\pi}{2}}\sqrt{1-\nu^{(i)}\sin^2\theta}\,d\theta,
	\ea
	\eeq
	and the corresponding analytical form of the Fourier coefficients are written as
	\beq
	\left.\ba{ll}\label{Fourier.coeff.amj}
	a_{j,0}^{(i)}&=\ds a_j^{(i)}+b_j^{(i)}\left(1-\f{1}{\nu^{(i)}}+\f{E(\nu^{(i)})}{\nu^{(i)} K^2(\nu^{(i)})}\right),\\
	a_{j,m}^{(i)}&=\ds \f{\pi^2b_j^{(i)}}{\nu^{(i)} K^2(\nu^{(i)})}\f{mq^{m}}{1-q^{2m}},~m=1,2,3,\dots,\\
	a_{j,-m}^{(i)}&=a_{j,m}^{(i)},~m=1,2,3,
	\ea\right\}(j,i)=(1,1),(2,1),(3,1),(1,3),
	\eeq
	and
	\beq
	\left.\ba{ll}\label{Fourier.coeff.amj.sncn}
	a_{j,0}^{(i)}&=0\\
	a_{j,m}^{(i)}&=\ds \f{-i\pi^2b_{j}^{(i)}}{\nu^{(i)} K^2(\nu^{(i)})}\sum\limits_{l=0}^{\infty}\left(\f{q^{2l+1}}{(1-q^{2l+2m+1})(1+q^{2l+1})}-\f{q^{2l+1}}{(1-q^{2l+1})(1+q^{2l+2m+1})}\right)\\
	&~+\ds\f{-i\pi^2b_{j}^{(i)}}{\nu^{(i)} K^2(\nu^{(i)})}\ds\sum\limits_{n_1,n_2=0,n_1+n_2+1=m}^{\infty}\f{q^{m}}{(1-q^{2n_1+1})(1+q^{2n_2+1})},~m=1,2,3,\dots\\
	a_{j,-m}^{(i)}&=-a_{j,m}^{(i)},~m=1,2,3,\dots,
	\ea\right\}(j,i)=(2,3),(3,3).
	\eeq
	Alternatively, we will apply the numerical methods for finding the Fourier coefficients for these two periodic solutions as well as for hyperbolic solutions. In result discussion section \ref{results}, we will compare analytical and numerical methods. There are few number of papers investigated the linear stability of the solutions of TWI \cite{kref96,xie,shen2018} with quadratic nonlinearity. It is noted that we cannot utilize the above-mentioned linear stability for constant wave (CW) solutions. So, in this special case, we will employ the plane wave expansion method for studying linear stability analysis \cite{a.p.and.p.a16.3,twri1st10,xie6}. \subsection{Modulation instability analysis}
	Modulation instability is essential for exploring and elucidating the emergence and development of large-amplitude periodic wave trains. Typically, this phenomenon is investigated when it arises as a consequence of the interaction between nonlinear and dispersive effects at an interface. The mechanism underlying modulation instability (MI) is responsible for generating solitons when nonlinear and dispersion coefficients engage in a confrontation, accompanied by the introduction of slight perturbations in a continuous wave (CW). The introduction of these minor perturbations results in the establishment of a linear expression, from which the dispersion relation of MI is derived. Numerous nonlinear systems exhibit instability, disrupting the steady-state of the system. 
	Now for the constant wave solutions (\ref{a5}) we have applied plane wave expansion method to study the modulation instability (MI) phenomenon.
	In this case we have considered the small perturbations to the solutions of the form
	\beq\label{MI}%\label{perturbation}
	F_j^{(5)}(t,z)=a_j^{(5)}\left(1+\delta P_j(t,z)\right)e^{i(\theta_j(t)+\mu_j z)},~~~j=1,2,3
	\eeq
	where $\delta\ll 1$ is infinitesimal amplitude of the perturbations and $P_j(t,z) (j=1,2,3)$ are all complex functions. We substitute Eq.(\ref{MI}) in Eq.(\ref{Eq.twi}) to linearize in $P_j(t,z) (j=1,2,3)$ and then we obtain the following equations
	\beq \label{Eq.MI}
	\ba{l}
	i\f{\partial P_1}{\partial z}-g_1\f{\partial^2 P_1}{\partial t^2}+\frac{\sigma_1 a_2 a_3}{a_1}\left(-P_1+P_2^*+P_3 \right)=0,\\
	i\f{\partial P_2}{\partial z}-g_2\f{\partial^2 P_2}{\partial t^2}+\frac{\sigma_2 a_1 a_3}{a_2}\left(-P_2+P_1^*+P_3 \right)=0,\\
	i\f{\partial P_3}{\partial z}-g_3\f{\partial^2 P_3}{\partial t^2}+\frac{\sigma_3 a_1 a_2}{a_3}\left(-P_3+P_1+P_2 \right)=0.\\
	\ea
	%\end{eqnarray}
	\eeq
	Now, we look at periodic perturbation of $P^*_j(t,z) (j=1,2,3)$ which diminish the differential equation Eq.(\ref{Eq.MI}) to a set of algebraic equations. So, we should adopt solution of Eq.(\ref{Eq.MI}) in the form
	\beq \label{FEq.MI}
	\ba{l}
	P_1(t,z)=f_1e^{i(\kappa t-\Omega z)}+f_2^*e^{-i(\kappa t-\Omega^* z)},\\
	P_2(t,z)=f_3e^{i(\kappa t-\Omega z)}+f_4^*e^{-i(\kappa t-\Omega^* z)},\\
	P_3(t,z)=f_5e^{i(\kappa t-\Omega z)}+f_6^*e^{-i(\kappa t-\Omega^* z)},
	\ea
	\eeq
	where the real parameter $\kappa$ denotes wave number (spatial frequency) and the frequency $\Omega$ determines instability which is a complex number and $f_k (k=1,2,3,4)$ are complex constants.\\
	We substitute Eq.(\ref{FEq.MI}) into Eq.(\ref{Eq.MI}) and obtain the following eigenvalue problem
	\beq\label{EV.MI}
	\left(\ba{cccccc}
	\frac{\sigma_1 a_2 a_3}{a_1}-g_1\kappa^2 & 0 & 0 & -\frac{\sigma_1 a_2 a_3}{a_1} & -\frac{\sigma_1 a_2 a_3}{a_1} & 0 \\
	0 & g_1\kappa^2-\frac{\sigma_1 a_2 a_3}{a_1} & \frac{\sigma_1 a_2 a_3}{a_1} & 0 & 0 & \frac{\sigma_1 a_2 a_3}{a_1} \\
	0 & -\frac{\sigma_2 a_1 a_3}{a_2} & \frac{\sigma_2 a_1 a_3}{a_2}-g_2\kappa^2 & 0 & -\frac{\sigma_2 a_1 a_3}{a_2} & 0 \\
	\frac{\sigma_2 a_1 a_3}{a_2} & 0 & 0 & g_2\kappa^2-\frac{\sigma_2 a_1 a_3}{a_2} & 0 & \frac{\sigma_2 a_1 a_3}{a_2} \\
	-\frac{\sigma_3 a_1 a_2}{a_3} & 0 & -\frac{\sigma_3 a_1 a_2}{a_3} & 0 & \frac{\sigma_3 a_1 a_2}{a_3}-g_3\kappa^2 & 0 \\
	0 & \frac{\sigma_3 a_1 a_2}{a_3} & 0 & \frac{\sigma_3 a_1 a_2}{a_3} & 0 & g_3\kappa^2-\frac{\sigma_3 a_1 a_2}{a_3} \\
	\ea\right)
	\left(\ba{c}f_1\\f_2\\f_3\\f_4\\f_5\\f_6\ea\right)=\Omega\left(\ba{c}f_1\\f_2\\f_3\\f_4\\f_5\\f_6\ea\right).
	\eeq
	%\subsection{MI}
	Since the matrix elements of the eigenvalue problem (\ref{EV.MI}) are real, therefore, it has six eigenvalues (real or complex) for each $\kappa$, say $\Om_i(i=1,2,\cdots,6)$. Therefore, the constant wave solution is unstable that is MI occurs if the system (\ref{EV.MI}) has a zero positive imaginary frequency ($\Omega$ act as the action of exponentially growth perturbation) for a fixed frequency $\kappa$. But the CW solution is always stable against the small perturbation growth if the system (\ref{EV.MI}) has only real frequencies. Thus, imaginary frequency is only necessary condition for existence of MI. To study the MI for CW solution we have defined the instability gain $G(\kappa)$ in the form \cite{twri1st10,malomed2}
	\beq\label{Gk}
	G(\kappa)=\max\limits_{i=1,2,\cdots,6} Im(\Om_i(\kappa)).
	\eeq
	The characteristic polynomial of the eigenvalue problem (\ref{EV.MI}) is defined by 
	\beq\label{char.pol}
	\Om^6+a_4\Om^4+a_2\Om^2+a_0=0,
	\eeq 
	where
	\beq
	\ba{ll}
	a_0&=4 a b c^2 x y - c^2 x^2 y^2 + 4 a b^2 c x z + 4 a^2 b c y z - 
	4 a b c x y z - 2 b c x^2 y z - 2 a c x y^2 z + 2 c x^2 y^2 z - 
	b^2 x^2 z^2 - 2 a b x y z^2 + 2 b x^2 y z^2\\
	&\quad  - a^2 y^2 z^2 + 
	2 a x y^2 z^2 - x^2 y^2 z^2\\
	a_2&=-4 a b c x + b^2 x^2 + 2 b c x^2 + c^2 x^2 - 4 a b c y + 2 a b x y - 
	2 b x^2 y + a^2 y^2 + 2 a c y^2 + c^2 y^2 - 2 a x y^2 + x^2 y^2 + 
	4 a b c z + 2 a c x z\\
	&\quad  - 2 c x^2 z + 2 b c y z - 2 c y^2 z + 
	a^2 z^2 - 2 a b z^2 + b^2 z^2 - 2 a x z^2 + x^2 z^2 - 2 b y z^2 + 
	y^2 z^2\\
	a_4&=-a^2 + 2 a b - b^2 - 2 a c - 2 b c - c^2 + 2 a x - x^2 + 
	2 b y - y^2 + 2 c z - z^2
	\ea
	\eeq 
	and
	\beq
	\ba{ll}
	a=\frac{\sigma_1 a_2 a_3}{a_1},x=g_1\kappa^2,\\
	b=\frac{\sigma_2 a_1 a_3}{a_2},y=g_2\kappa^2,\\
	c=\frac{\sigma_3 a_1 a_2}{a_3},z=g_3\kappa^2.
	\ea
	\eeq
	Therefore, the characteristic roots of the polynomial (\ref{char.pol}) are obtained from the cubic equation
	\beq\label{cardan}
	q^3+3b_1q+b_0=0, 	
	\eeq 
	where $q=\ds\Om^2+\frac{a_4}{3}$, $b_1 =\ds \frac{a_2}{3} - \frac{a_4^2}{9}$, $b_0 =\ds a_0 -\frac{a_2 a_4}{3}$.
	Then we can find the roots $q_1,q_2,q_3$ of the cubic equation (\ref{cardan}) by Cardan's method (for details see the Appendix). 
	Finally, we obtain the required six eigenvalues by solving three simple quadratic equations  
	\beq
	\ba{ll}
	\Om^2&=q_i-\frac{a_4}{3},~i=1,2,3.
	\ea 
	\eeq  
	For each $\kappa$, the maximum absolute value of imaginary parts among six characteristic roots are defined by the instability gain $G(\kappa$).
	The MI of CW solutions to the second-harmonic-generation in a multidimensional dispersive medium investigated analytically in absence of second order dispersion \cite{malomed2}.
	
	\section{Results and discussions}\label{results}
	In this section, we investigate the influence of group velocity dispersion and wave vector mismatch on stability of two sets of self-consistent periodic solutions expressed by Jacobi elliptic functions; two sets of one hump, two hump bright soliton solutions and a set of constant magnitude wave solution. Also, we explore the linear stability analysis by the nature of $\lambda$ and propagation dynamics for diverse parameters. To find the values of $\lambda$ of the eigenvalue equation (\ref{Eq.EV.cne0}) for a given solution $F_j^{(i)}(j=1,2,3)$, we have applied the Fourier collocation method (FCM) \cite{pla} and Finite difference method (FDM) \cite{ND}. In FCM, the Fourier coefficients are defined analytically as well as numerically for self-consistent periodic solutions and numerically for soliton solutions. The solution $F_j^{(i)}(j=1,2,3)$ is unstable if perturbed eigen function increases with perturbation growth rate $\lam_R>0$ during propagation, where $\lambda=\lambda_R+i\lambda_I$. In this case, perturbation is the eigenvector of (\ref{Eq.EV.cne0}) corresponds to the eigenvalue with largest positive real part $\lambda_R$. For a given solution $F_j^{(i)}(j=1,2,3)$, if $\lambda_R\le 0$ then we have added $5\%$ random noise to the initial solution as a perturbation. On the other hand, the intensity profiles of solutions are verified via propagation dynamics. Now for propagation dynamics, we have added the above mentioned perturbations to the initial solution and then apply (i) Pseudospectral method for second-order spatial derivatives in the direction of $t$ and $4$th order Runge-Kutta (RK$4$) method for the temporal derivative \cite{pla} in the direction of $z$ and (ii) second-order central-difference formula for second-order spatial derivatives in the direction of $t$ and Crank-Nicholson finite difference method for the temporal derivative \cite{ND,Crank-Nicolson,Crank-Nicolsonr2} in the direction of $z$ with the step lengths $dt=0.01$ and $dz=0.0005$. In order to obtain the intensity profile of TWI whose amplitudes are defined in Eqs. (\ref{ansatz1}), (\ref{ansatz1.nu1}), (\ref{ansatz2A1}) and (\ref{Ansatz2.A1.nu1}), we have made all the special grid points as $t_j=-L+\delta t_j$, $j=1,2,3,...,N+1$ ($L$ being the half-width), where $\delta t=2L/N$ is taken to be the grid dividing (resolution) and both of the boundary points are denoted by $j=1$(left) and $j=N+1$(right). 
	\begin{figure}[h] % 3
		\centering	
		\includegraphics[width=6.5cm,height=5.5cm]{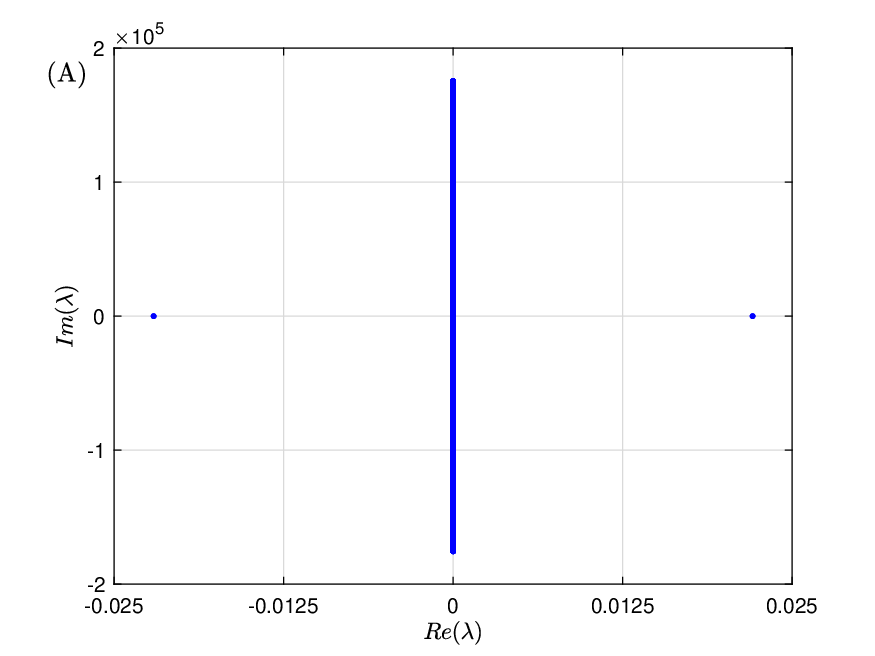}~
		\includegraphics[width=6.5cm,height=5.5cm]{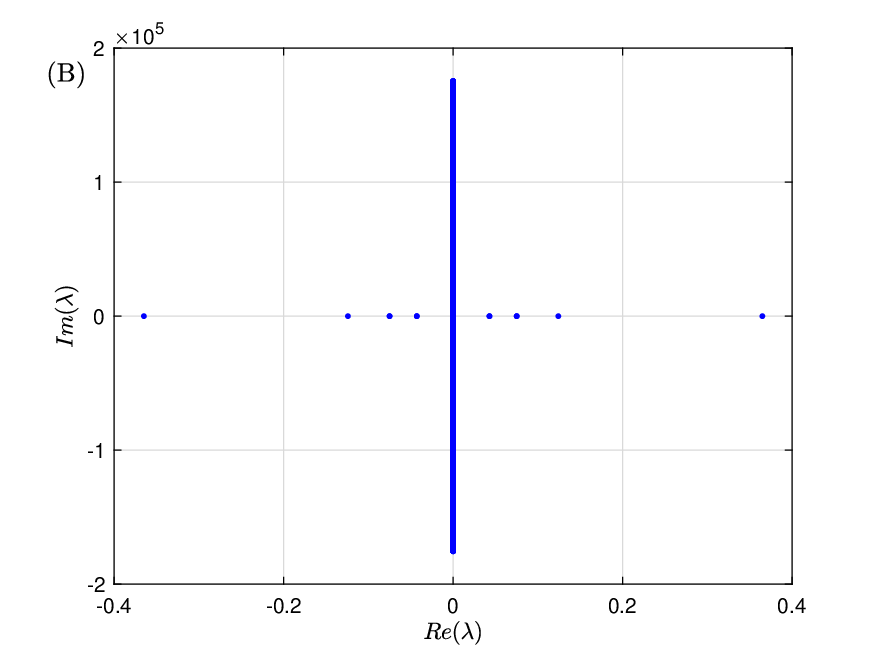}\\
		\includegraphics[width=6.5cm,height=6cm]{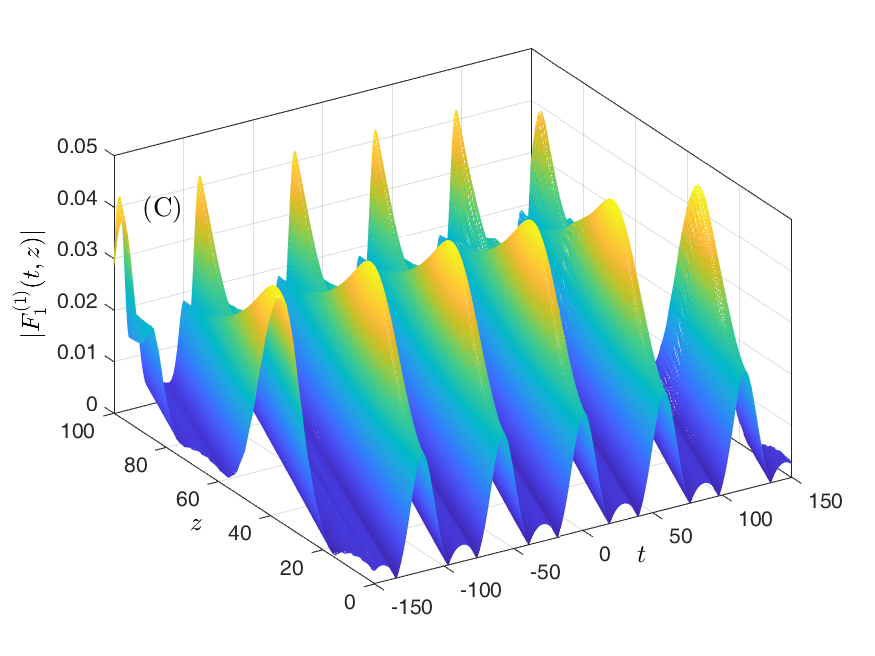}~
		\includegraphics[width=6.5cm,height=6cm]{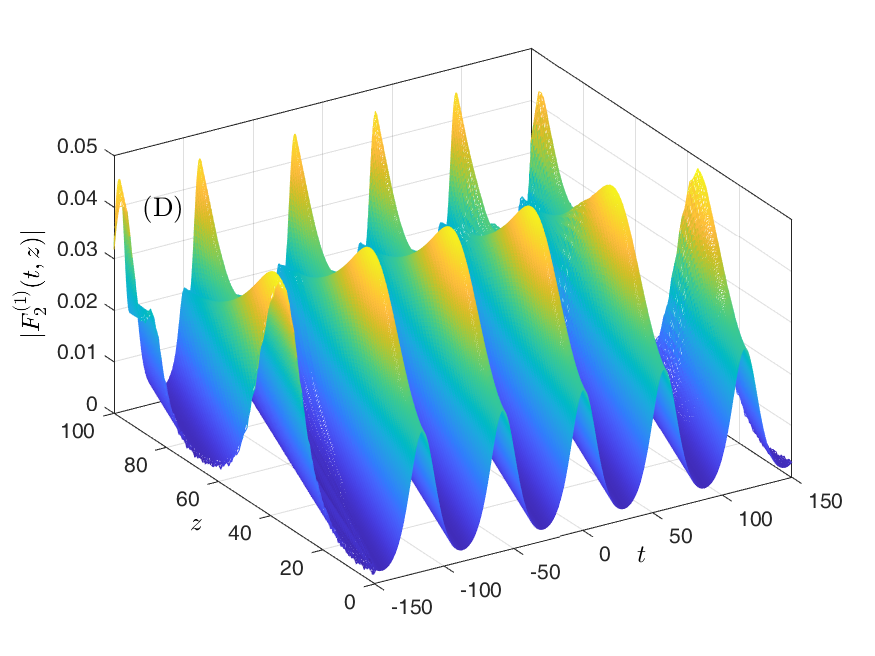}~
		\includegraphics[width=6.5cm,height=6cm]{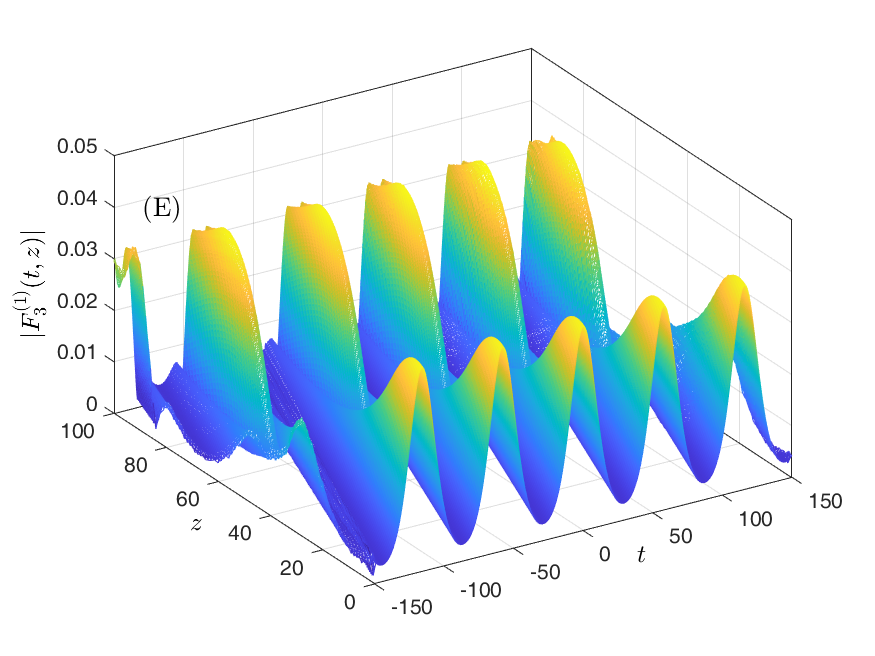}
		\caption{\label{fig3} Plot of the unstable mode of Eqs.(\ref{F=Aetheta}) corresponding to (\ref{ansatz1}). The parameters are $\nu^{(1)}=0.85, \Delta k=-0.04,\omega_1=\omega_2=\omega_3=1, \sigma=1, v_1=v_2=v_3=1, g_1=1, g_2=1.1$. In (A) eigenvalues with analytical Fourier coefficients and in (B) eigenvalues with numerical Fourier coefficients are plotted and the corresponding propagation dynamics of $F_j^{(1)}$ for $j=1,2,3$ are compared in (C) -(E) respectively. },
	\end{figure}
	\begin{figure}[h] % 4
		\centering	
		\includegraphics[width=6.5cm,height=5.5cm]{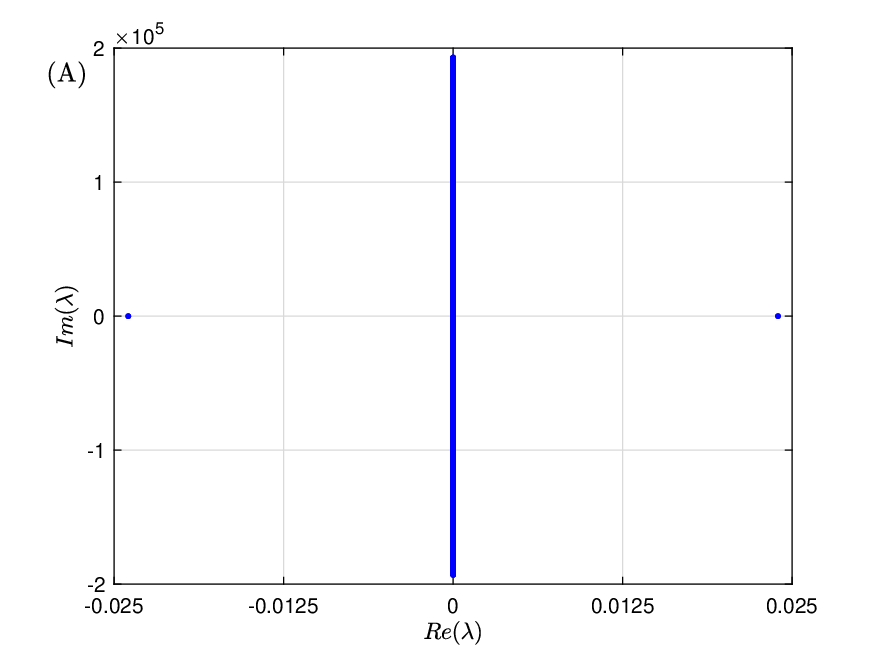}\\
		\includegraphics[width=6.5cm,height=6cm]{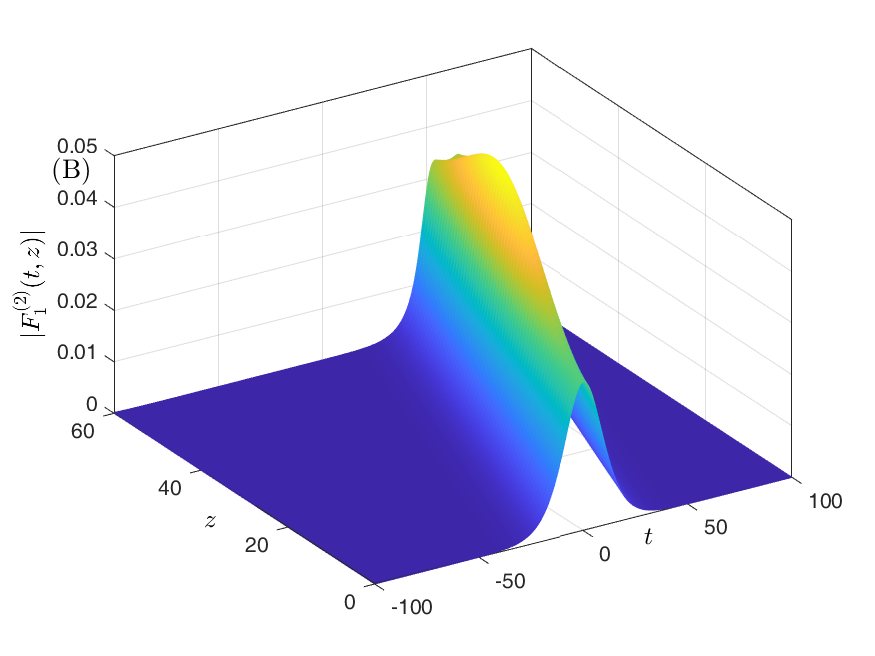}~
		\includegraphics[width=6.5cm,height=6cm]{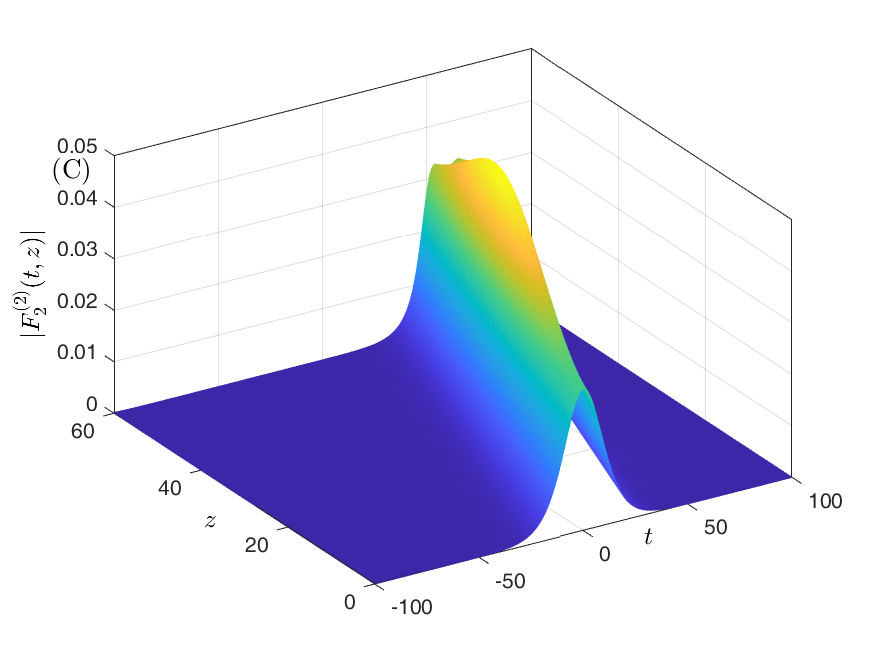}~	
		\includegraphics[width=6.5cm,height=6cm]{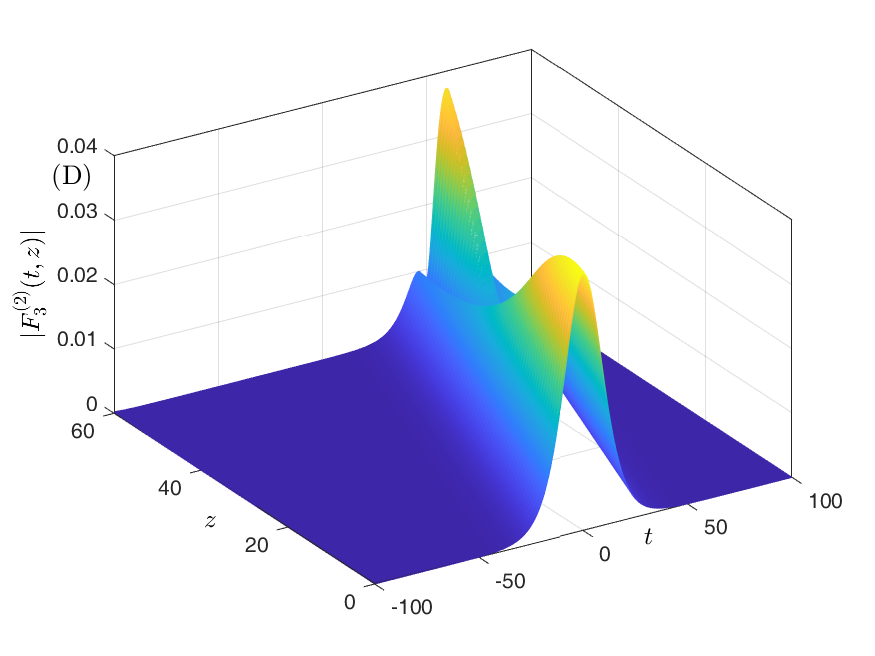}
		\caption{\label{fig4} Plot of unstable mode of Eqs.(\ref{F=Aetheta}) corresponding to (\ref{ansatz1.nu1}). The parameters are $\nu^{(2)}=1, \Delta k=-0.04,\omega_1=\omega_2=\omega_3=1, \sigma=1, v_1=v_2=v_3=1, g_1=1, g_2=1.1$. Fig. (A) shows the eigenvalues and Figs. (B)-(D) show comparison of propagation dynamics of $F_j^{(2)}$ for $j=1,2,3$.}
	\end{figure}
	%%%%%%%%%%%%%%%%%%%%%%%%%%%%%%%%
	\subsection{Stability of periodic solutions (\ref{ansatz1})}
	To validate, the linear stability of Jacobi elliptic solutions (8) for ansatz I, we have applied numerical technique as well as analytical Fourier coefficients. In case of analytical method, we have considered the Fourier coefficients from (\ref{Fourier.coeff.amj}) to find the maximum value of real parts of $\lambda$. In Fig. 3 (A), (B) we have plotted the eigenvalues of for the solutions (8) analytically and numerically respectively, for a particular set of parameters $\Delta k=-0.04,\omega_1=\omega_2=\omega_3=1, \sigma=1, v_1=v_2=v_3=1$ with the group velocity dispersion $g_1=1, g_2=1.1$ and $\nu^{(1)}=0.85$ which satisfy Eq. (\ref{muFirst}). As we have mentioned earlier that the given solution is unstable if the eigen-frequencies $\lambda$ have a positive real part. Now, for ansatz I, we have checked that the eigen frequencies have positive real parts $(\lambda_R)$ for any values of parameters.  Therefore, the solution (8) is always unstable for $0<\nu^{(1)}<1$. 
	For a careful study of numerical simulation of TWI, we have considered the following boundary conditions 
	\beq
	\left.\ba{ll}
	F_j^{(1)}(0,z)=\left(a_j^{(1)}+b_j^{(1)}\right)e^{i\mu_jz},\\
	F_j^{(1)}(\f{2K(\nu^{(1)})}{\alpha^{(1)}},z)=\left(a_j^{(1)}+b_j^{(1)}\right)e^{i\left(\frac{2K(\nu^{(1)})t}{2g_jv_j\alpha^{(1)}}+\mu_jz\right)},\\
	F_j^{(1)}(t,0)=A_j^{(1)}(t)e^{\frac{it}{2g_jv_j}},
	\ea\right\},j=1,2,3.
	\eeq
	To visualize dynamics of the periodic solution we have calculated the numerical values of TWI in the domain $t\in[-L,L],z\in[0,100]$. The initial solutions of each components develop five humps in $t\in[-L,L]$, $L=150$. The propagation dynamics are shown in Fig. \ref{fig3} (C)-(E). From this figure one can observe that the first two components are almost similar. Each intensity profile dramatically changes the magnitude and shape with as $z$ increases. Therefore, the self-periodic solution of the first set is unstable. 
	%%%%%%%%%%%%%%%%	
	\subsection{Stability of one hump bright soliton solutions (\ref{ansatz1.nu1})}
	For the one hump bright soliton, the amplitude of TWI in solution (\ref{ansatz1.nu1}) is never zero but rapidly goes to a finite number at infinity, that is 
	\beq
	\left.\ba{ll}
	|F_j^{(2)}(t,z)|\rightarrow|a_j^{(2)}|,|t|\rightarrow\infty,\\
	F_j^{(2)}(t,0)=A_j^{(2)}(t)e^{\frac{it}{2g_jv_j}}
	\ea\right\},j=1,2,3.
	\eeq 
	In this case we have considered the following boundary conditions  
	\beq
	\left.\ba{ll}
	F_j^{(2)}(-L,z)=a_j^{(2)}e^{i\left(\frac{-L}{2g_jv_j}+\mu_jz\right)},\\
	F_j^{(2)}(L,z)=a_j^{(2)}e^{i\left(\frac{L}{2g_jv_j}+\mu_jz\right)},\\
	F_j^{(2)}(t,0)=A_j^{(2)}(t)e^{\frac{it}{2g_jv_j}}
	\ea\right\},j=1,2,3.
	\eeq 
	The eigenvalue equation (\ref{Eq.EV.c0}) is solved numerically by FCM in presence of $F_j^{(2)},j=1,2,3$. In particular, Fig. \ref{fig4} (A) shows the numerical eigenvalues of an unstable mode for the parameters $\Delta k=-0.04,\omega_1=\omega_2=\omega_3=1, \sigma=1, v_1=v_2=v_3=1$ with the group velocity dispersion $g_1=1, g_2=1.1$. Under the same parameters, propagation dynamics of three component bright solitons are shown in Fig. \ref{fig4} (B)-(D). The intensity profile of each component is changing the shape drastically as $z$ changes. From Fig. \ref{fig4}, one can observe that, three waves represent unstable one hump bright soliton solutions. An important result we noted that one hump bright soliton solution (\ref{ansatz1.nu1}) always unstable. 
	%%%%%%%%%%%%%%%%%%%%%%%%%%%%%%%%%%%%%%%%
	\subsection{Stability of periodic solutions (\ref{ansatz2A1})}
\begin{figure}[ht] % 5
		\centering
		\includegraphics[width=6.5cm,height=5.5cm]{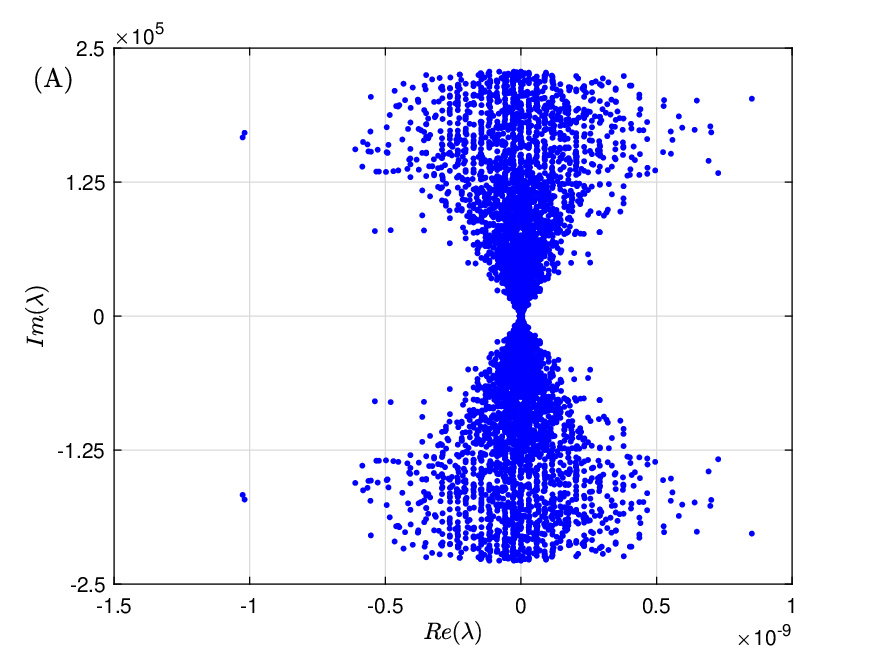}~
		\includegraphics[width=6.5cm,height=5.5cm]{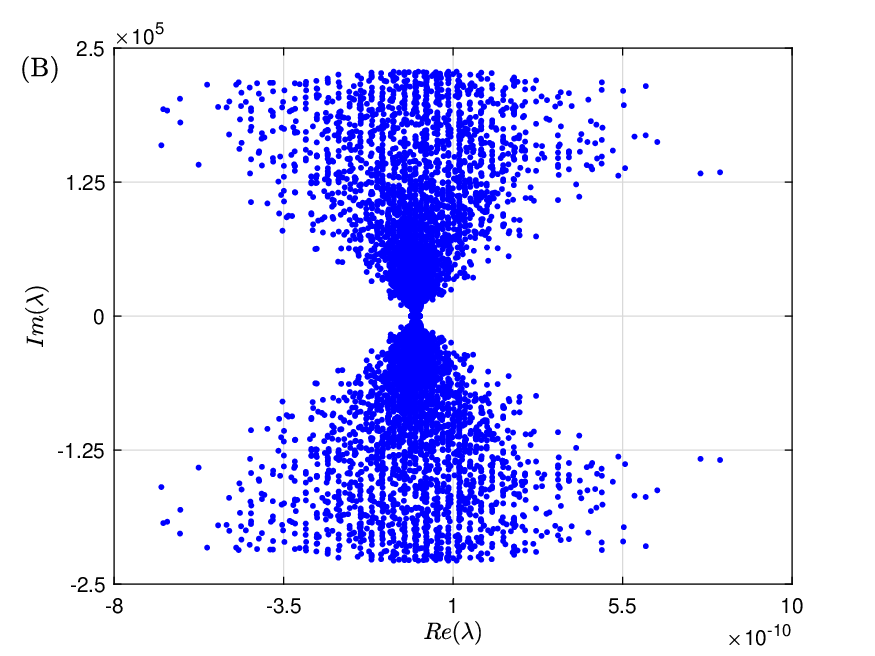}\\	
		\includegraphics[width=6.5cm,height=6cm]{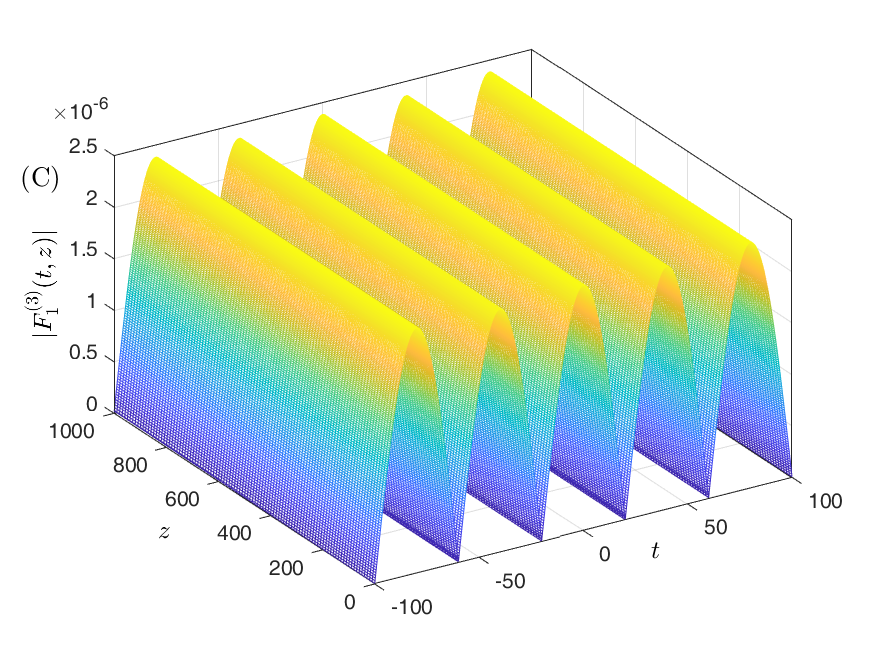}~
		\includegraphics[width=6.5cm,height=6cm]{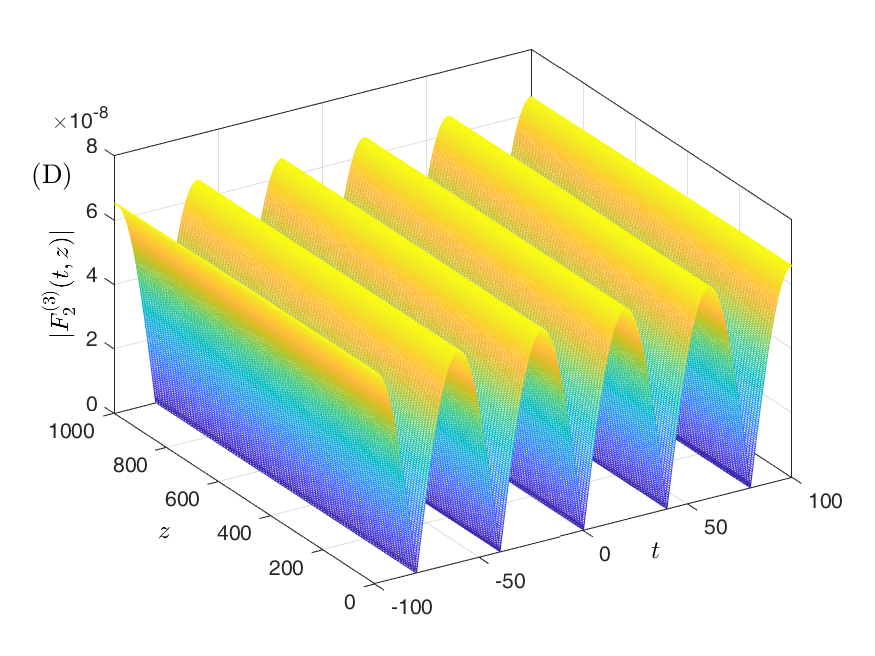}~
		\includegraphics[width=6.5cm,height=6cm]{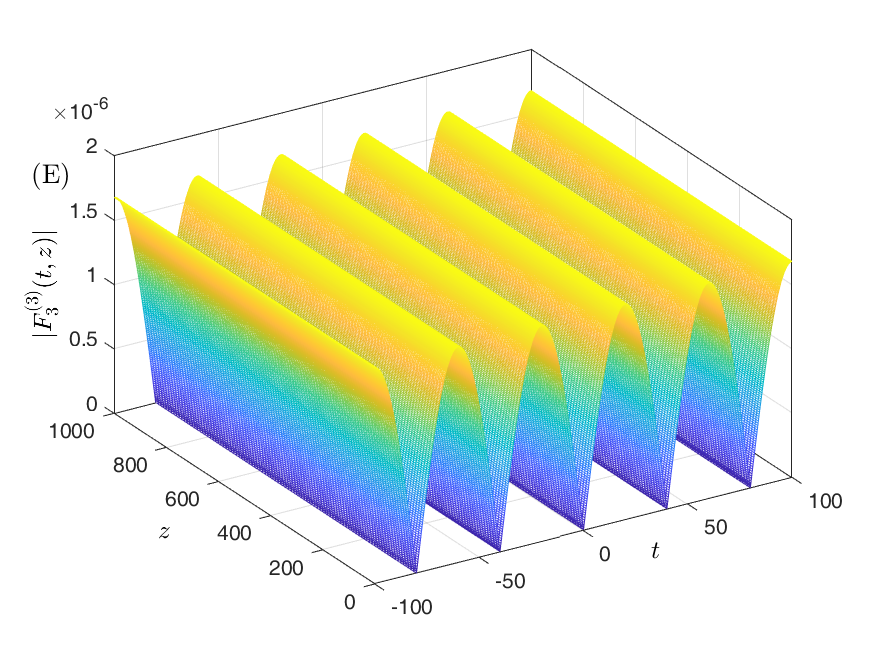}		
		\caption{\label{fig5} Plot of the stable mode of Eq.(\ref{F=Aetheta}) corresponding to (\ref{ansatz2A1}). The parameters are $\nu^{(3)}=0.01, \Delta k=0.008, \omega_1=\omega_2=\omega_3=1, \sigma=1, v_1=-2\times10^8, v_2=v_3=10^8, g_1=-0.001, g_2=1.3$. Figs. (A) and (B) show the eigenvalues corresponding to numerical and analytical Fourier coefficients respectively. The propagation dynamics of $F_j^{(3)}$ for $j=1,2,3$ are compared in Fig. (C)-(E) with $5\%$ random noise.},
	\end{figure}

	Similarly, for ansatz II, in case of analytical method, we have considered the Fourier coefficients from (\ref{Fourier.coeff.amj}) and (\ref{Fourier.coeff.amj.sncn}) to find the largest real part of $\lambda$. The eigenvalues are shown in Fig. \ref{fig5}(A), (B) using analytical and numerical Fourier coefficients respectively for a particular values of parameters $\Delta k=0.008, \omega_1=\omega_2=\omega_3=1, \sigma=1, v_1=-2\times10^8, v_2=v_3=10^8, g_1=-0.001, g_2=1.3$ and $\nu^{(3)}=0.01$. We have found that real parts of $\lam$s' are too small and they lie between $-8\times10^{-10}$ to $10^{-9}$. Therefore, we can say that the solution (18) is stable.
	Similarly, for propagation dynamics of TWI we have considered the following boundary conditions 
	\beq
	\ba{ll}
	\left.
	\ba{ll}F_1^{(3)}(0,z)=\left(a_1^{(3)}+b_1^{(3)}\right)e^{i\mu_1z},\\
	F_1^{(3)}(\f{4K(\nu^{(3)})}{\alpha^{(3)}},z)=\left(a_1^{(3)}+b_1^{(3)}\right)e^{i\left(\frac{4K(\nu^{(3)})t}{2g_1v_1\alpha^{(3)}}+\mu_1z\right)},
	\ea 
	\right.\\
	\left.
	\ba{ll}F_j^{(3)}(0,z)=a_j^{(3)}e^{i\mu_jz},\\
	F_j^{(3)}(\f{4K(\nu^{(3)})}{\alpha^{(3)}},z)=a_j^{(3)}e^{i\left(\frac{4K(\nu^{(3)})t}{2g_jv_j\alpha^{(3)}}+\mu_jz\right)},
	\ea
	\right\},j=2,3,\\
	\left.
	\ba{ll} 
	F_j^{(3)}(t,0)=A_j^{(3)}(t)e^{\frac{it}{2g_jv_j}}
	\ea\right.,j=1,2,3,
	\ea 
	\eeq
	and the corresponding stable propagation dynamics are shown in Fig. \ref{fig5}(C) -(E). Due to the boundary conditions, propagation dynamics of the second and third components are same in shape but their magnitudes are different. One can observe that for large values of group velocities the intensity profiles are very low and from Fig. \ref{fig5}, it is clear that $|F_1^{(3)}(t,z)|\le 2.5\times10^{-6}$, $|F_2^{(3)}(t,z)|\le 8.0\times10^{-8}$ and $|F_3^{(3)}(t,z)|\le 2.0\times10^{-6}$.
	%%%%%%%%%%%%%%%%%%%%%%%%%%%%%%%%%%%%%%%%%%%%%%%%%%%%%%%%%%%%%
	\subsection{Stability of one, two hump bright soliton solutions (\ref{Ansatz2.A1.nu1})}
\begin{figure}[ht] % 6
		\centering	
		\includegraphics[width=6.5cm,height=5.5cm]{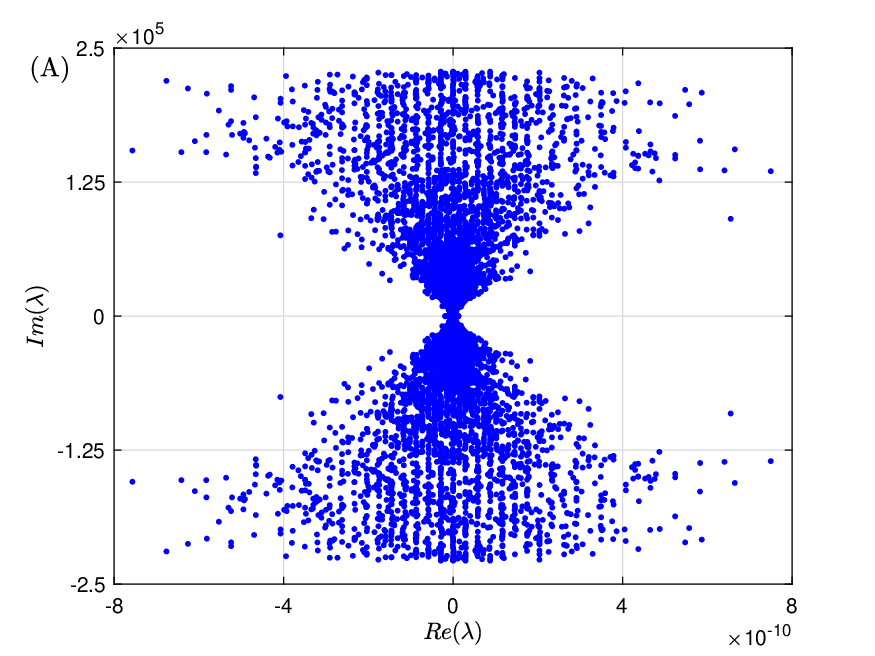}\\
		\includegraphics[width=6.5cm,height=6cm]{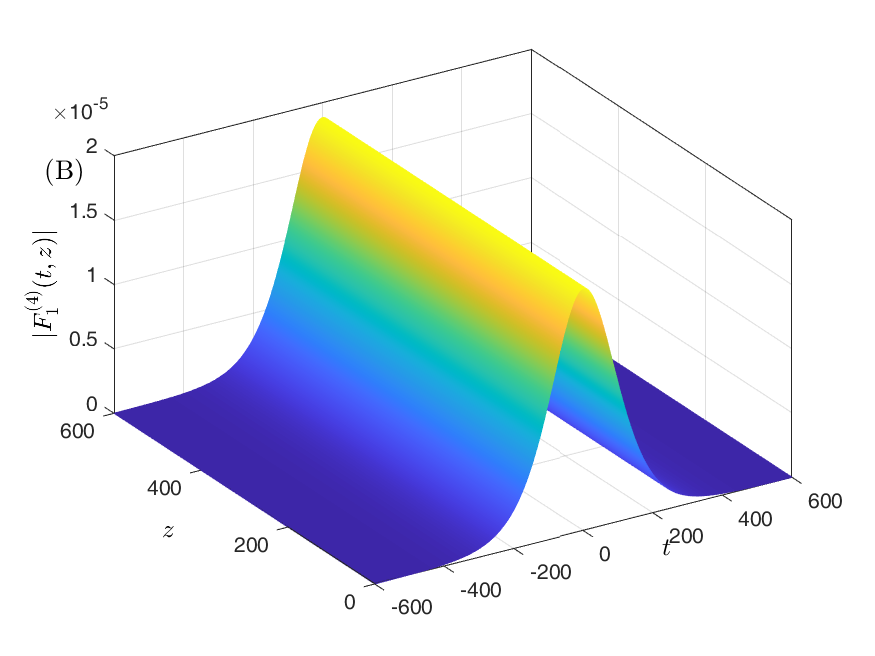}~
		\includegraphics[width=6.5cm,height=6cm]{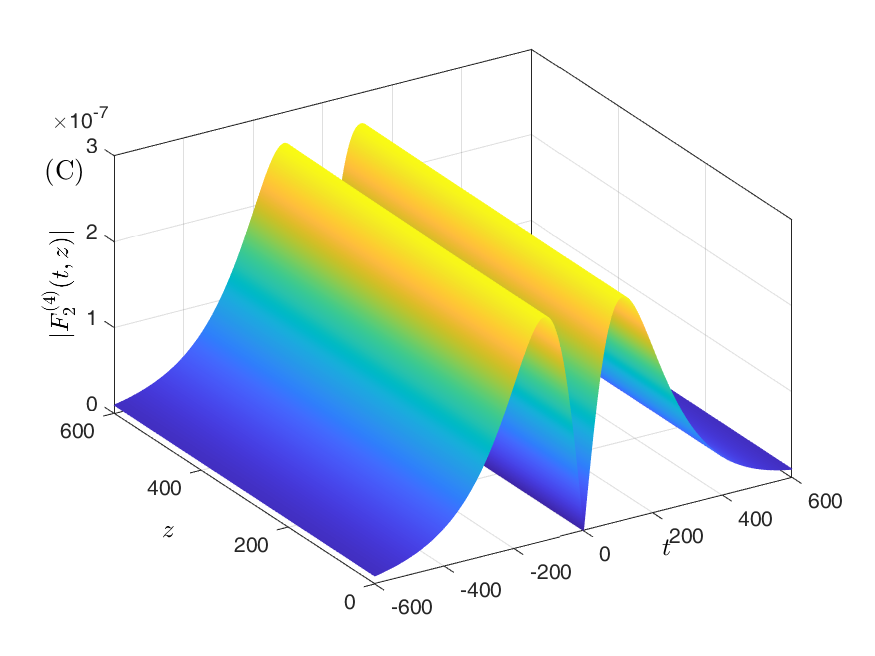}~	
		\includegraphics[width=6.5cm,height=6cm]{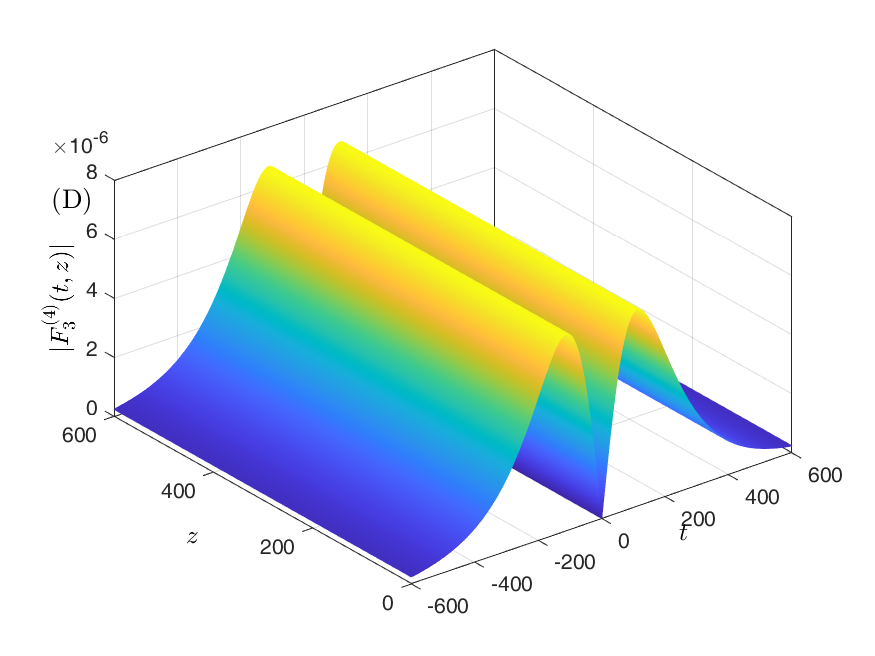}
		\caption{\label{fig6} Plot of the stable mode of Eqs.(\ref{F=Aetheta}) corresponding to (\ref{Ansatz2.A1.nu1}). The parameters are $\nu^{(4)}=1, \Delta k=-0.00008$ and other parameters are taken from Fig. \ref{fig5}.  In (A) eigenvalues of unstable mode of one and two hump bright solitons are plotted and their dynamics of $F_j^{(4)}$ for $j=1,2,3$ are shown in Fig. (B)-(D) with $5\%$ random noise.}
	\end{figure}

	Due to the presence of $sech\alpha^{(4)} t\tanh\alpha^{(4)} t$ term in the amplitude of $F_j^{(4)},~j=2,3$, we obtain two hump bright solitons. Again the amplitude of TWI in solution (\ref{Ansatz2.A1.nu1}) is never zero but rapidly goes to zero at infinity, that is 
	\beq
	\left.\ba{ll}
	|F_j^{(4)}(t,z)|\rightarrow0,|t|\rightarrow\infty,\\
	F_j^{(4)}(t,0)=A_j^{(4)}(t)e^{\frac{it}{2g_jv_j}}
	\ea\right\},j=1,2,3,
	\eeq 
	which is different from one hump bright soliton (\ref{ansatz1.nu1}). In this case we have considered the following boundary conditions
	\beq
	\left.\ba{ll}
	F_j^{(4)}(-L,z)=F_j^{(4)}(L,z)=0,,\\
	F_j^{(4)}(t,0)=A_j^{(4)}(t)e^{\frac{it}{2g_jv_j}}
	\ea\right\},j=1,2,3.
	\eeq 
	To visualize, the linear stability, eigenvalues are plotted in Fig. \ref{fig6} (A) for soliton solutions with wave vector mismatch $\Delta k=-0.00008, \omega_1=\omega_2=\omega_3=1, \sigma=1$, group velocity $v_1=-2\times10^8, v_2=v_3=10^8$, group velocity dispersion $g_1=-0.001, g_2=1.3$, $\nu^{(4)}=1$. From this figure, one can observe that the absolute values of real parts of $\lambda$s' are less than $8\times10^{-10}$. For the same values of the parameters, propagation of components of TWI are plotted in Fig. \ref{fig6}(B)-(D). It is clear from Fig. \ref{fig6} (B) that first component of (\ref{Ansatz2.A1.nu1}) is a stable one hump bright soliton and from Fig. \ref{fig6} (C), (D) one can say that the second, third components are stable two hump bright solitons. Again one can observe that intensity profiles of stable propagation are low such that $|F_1^{(4)}(t,z)|\le 2.0\times10^{-5}$, $|F_2^{(4)}(t,z)|\le 3.0\times10^{-7}$ and $|F_3^{(4)}(t,z)|\le 8.0\times10^{-6}$.  
	
	%%%%%%%%%%%%%%%%%%%%%%%%%%%%%%%%%%%%%%%%%%%%%%%%%%%%%%%%%%%%%%%%%%%%%%%%%%%%%%
	%%%%%%%%%%%%%%%%%%%%%%%%%%%%%%%%%%%%%%%%%%%%%%%%%%%%%%%%%%%%%%%%%%%%%%%%%%%%%%		
	\subsection{Stability of constant magnitude wave solution (\ref{a5})}	
\begin{figure}[ht] % 8
		\centering	
		\includegraphics[width=14cm,height=8cm]{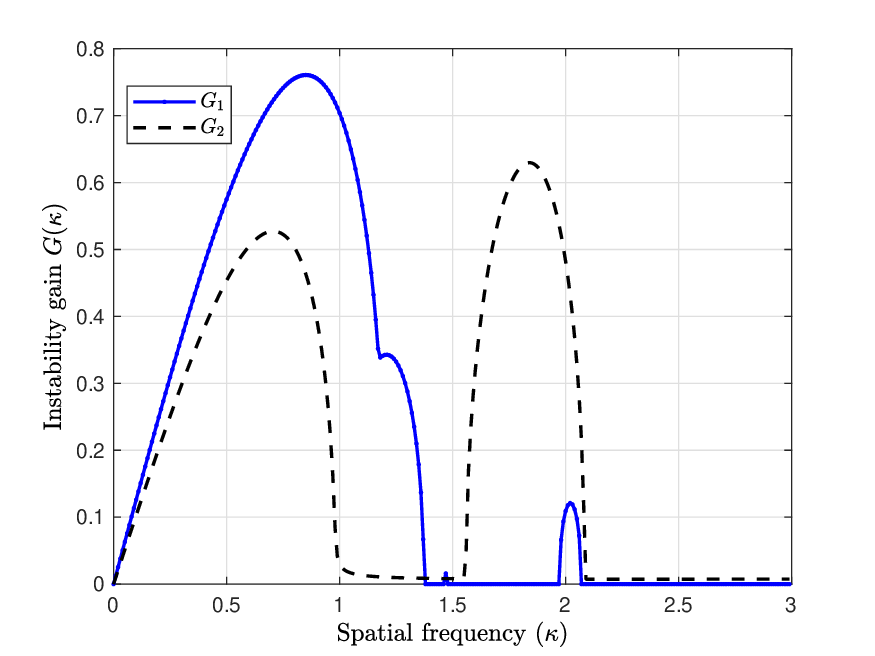}
		\caption{\label{fig7} Instability gain for $\Delta k=-0.04$. To draw the curves $G_1$, $G_2$ we have considered the values of $\om_1,\om_2,\om_3,\sigma,v_1,v_2,v_3$ $g_1,g_2$ from Fig. \ref{fig4}, \ref{fig6} respectively and $\mu_1^{(5)}=-1=\mu_2^{(5)}$.}
	\end{figure}

	In this section, we have applied the standard numerical strategies to calculate the instability gain $G(\kappa)$ for each $\kappa$. To understand MI of CW solution, we have plotted the instability gains versus the spatial frequency $\kappa$ in Fig. \ref{fig7} for two sets of parameters with a fixed wave vector mismatch $\Delta k=-0.04$ in presence of group velocity dispersion. To draw the MI gain curves $G_1$, $G_2$ we have taken values of $\om_1,\om_2,\om_3,\sigma,v_1,v_2,v_3$ $g_1,g_2$ from Figs. \ref{fig4}, \ref{fig6} respectively and $\mu_1^{(5)}=-1$, $\mu_2^{(5)}=-1$. From Fig. \ref{fig7}, it is observed that each MI gain curve has a maximum value and the CW solution is unstable. The varying instability gains for different set of parametric values ensures the instability of CW solutions. In ref. \cite{twri1st10}, MI has been compared with and without a group velocity dispersion. One can compare these results with the ref. \cite{twri1st10}. The MI was demonstrated experimentally in a ref. \cite{k97.7} and also it has been studied in plasma, fluid dynamics and optics.    
	
	%%%%%%%%%%%%%%%%%%%%%%%%%%%%%%%%%%%%%%%%%%%%%%%%%%%%%%%%%%%%%%%%%%%%%%%%%%%%%%
	
\begin{figure}[ht] % 6
		\centering	
		\includegraphics[width=6.5cm,height=6cm]{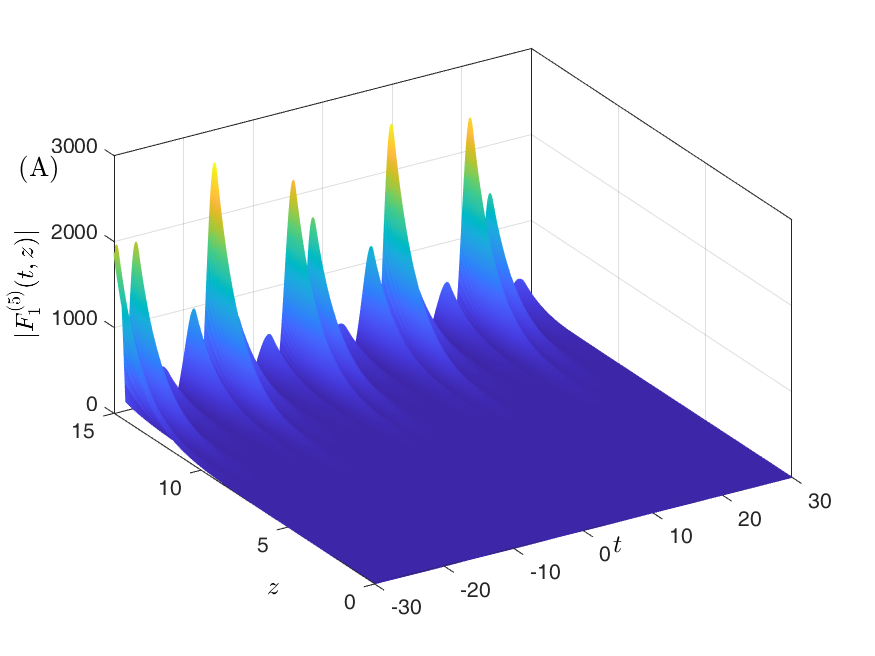}~
		\includegraphics[width=6.5cm,height=6cm]{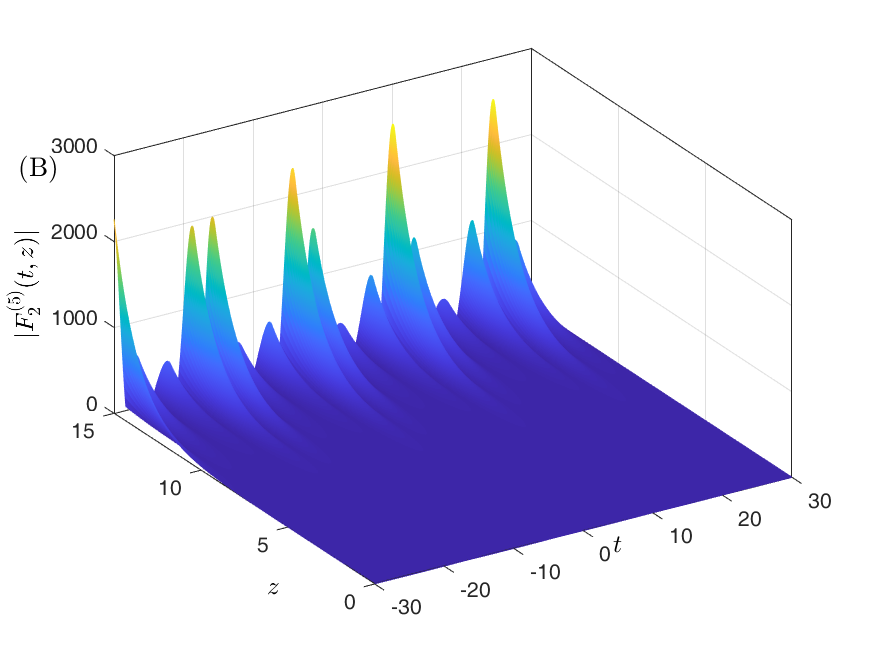}~
		\includegraphics[width=6.5cm,height=6cm]{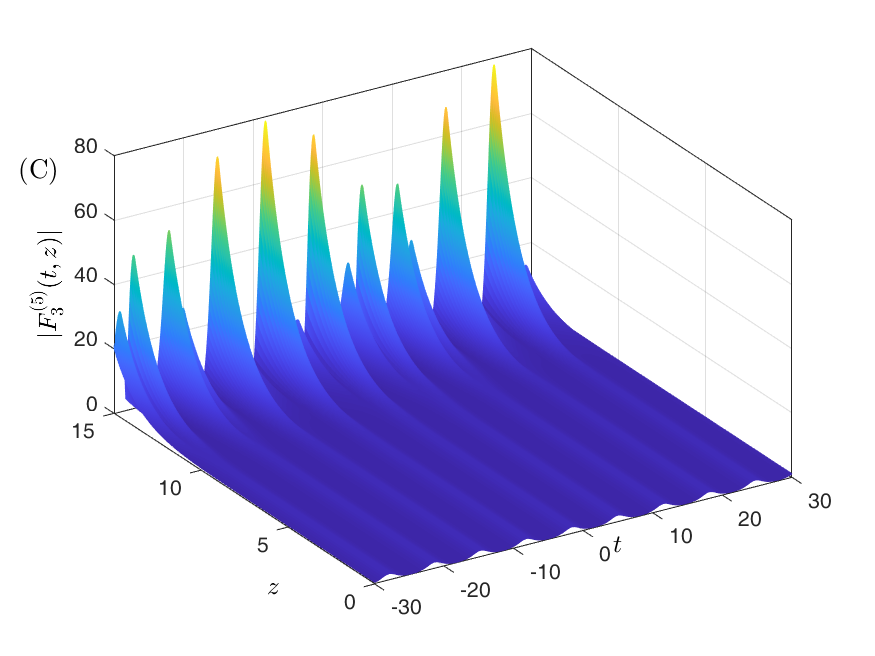}~	
		\caption{\label{fig8} Plot of dynamic evolution of intensities for $\Delta k=-0.04$, $v_1=v_2=v_3=1$, $g_1=g_2=1.1$, $\mu_1=\mu_2=-1$, $\omega_1=\omega_2=\omega_3=\sigma=1$. The perturbation parameters are $\max Im(\Om)=0.7605$ for $\kappa=0.861836$ and $\delta=0.01$.}
	\end{figure}

	We observe that if $b_0^2+4b_1^3=0$, then the most unstable eigenmode occurs. The most unstable eigenmode is defined numerically as $\left(f_1, f_2, f_3, f_4, f_5, f_6\right)=(0.5026 + 0.0000i, -0.0061 + 0.5025i, -0.4967 - 0.0180i, -0.0120 - 0.4968i, 0.0168 - 0.0127i, -0.0129 + 0.0167i)$ to the corresponding maximum $Im(\Om)=0.7605$ with spatial frequency $\kappa=0.861836$ for $\Delta k=-0.04$, $v_1=1, v_2=1, v_3=1$, $g_1=1, g_2=1.1$, $\mu_1=-1, \mu_2=-1$, $\omega_1=1,\omega_2=1, \omega_3=1, \sigma=1$. The nonlinear propagation dynamics of spatial profiles $F_j^{(5)}$, $j=1,2,3$ which are defined in Eq.(\ref{MI}) are presented in Fig. \ref{fig8} (A)-(C) in presence of most unstable periodic plane wave perturbations where maximum $Im(\Om)=0.7605$ at $\kappa=0.861836$ and $\delta=0.01$. Form Fig. \ref{fig8}, one can say that intensity profile of each component of TWI increases exponentially along the propagation direction $z$. 
	
	%%%%%%%%%%%%%%%%%%%%%%%%%%%%%%%%%%%%%%%%%%%%%%%%%%%%%%%%%%%%%%%%%%%%%%%%%%%%%%
	As a result, we can say that the solutions in first (\ref{ansatz1}) and second (\ref{ansatz1.nu1}) sets are always unstable for any parameters but the solutions in third (\ref{ansatz2A1}) and fourth (\ref{Ansatz2.A1.nu1}) sets have stable modes for some particular values of parameters. Also, we conclude that CW solution (\ref{a5}) is always unstable.
	
	\section{Conclusion}\label{conclusion}
	In this paper, we have investigated the existence of periodic and non-periodic solutions of three wave interacting system with group velocity dispersion and wave number mismatch in a quadratic nonlinear medium. We have obtained five sets of solutions for the first time to our knowledge. Among these sets (a) two sets contain periodic waves; (b) two sets have soliton waves and (c) one set of constant wave solutions. Functional relations among the parameters have been shown for various solutions. Then investigate the linear stability analysis of five sets of solutions. For periodic solutions, Fourier coefficients have been defined numerically as well as analytically for investigating the linear stability analysis. We have noticed that solutions of the first and second sets are always unstable for any kind of wave frequencies, group velocities, group-velocity dispersion, nonlinear coupling constant and wave vector mismatch. It is to be noted that solutions of third and fourth sets are stable for small values of wave vector mismatch ($\Delta k$). For large values of group velocities intensity profile of each component of TWI is too low. Moreover, we have employed the MI to analyse the linear stability of constant wave solutions. The instability gain spectrum and its variation of CW solutions have been reported for three sets of parameters. Our model can be applied for the nonlinear optical lattices with $\chi^{(2)}$-type non-linearity. Moreover, this proposed mathematical idea is a new approach in wave interacting system and can be applied to theoretical examples of other nonlinear dynamics and would also be used in experimental study which may open new research possibility in the field of nonlinear dynamics.
	%%%%%%%%%%%%%%%%%%%%%%%%%%%%%%%%%%%%%%%%%%%%%%%%%%%%%%%%%%%%%%%%%%%%%%%%%%%%%%%%%%%%%%%%%%%%%%%%%%%%%%%%%%%%%%%%%%%%%%%%%%%%%%%%%%%%%%%%%%%%%%%%%%%%%%%%
	\section*{Acknowledgment}
	DN dedicates this article to the loving of his kind hearted teacher, late Prof. Arun Kumar Chatterjee, Bose Institute. He gratefully acknowledges financial support from TARE, DST-SERB, New Delhi (sanction order: TAR/2021/000142). AD gratefully acknowledges the financial support under FIST, DST (letter number: SR/FST/MS-I/$2019/42$) to the Department of Mathematics, University of Kalyani. AD was also supported by DST-SERB (file number: EEQ/2022/000719). NG acknowledges financial support from SVMCM, Govt. of West Bengal.
	%%%%%%%%%%%%%%%%%%%%%%%%%%%%%%%%%%%%%%%%%%%%%%%%%%%%%%%%%%%%%%%%%%%%%%%%%%%%%%%%%%%%%%%%%%%%%%%%%%%%%%%%%%%%%%%%%%%%%%%%%%%%%%%%%%%%%%%%%%%%%%%%%%%%%%%%%
	\section*{Conflict of interest}
	The authors declare that they have no known competing financial interests.
	\section*{Data availability}
	All data generated or analyzed during this study are included in this article.
	\section*{Author contributions}
	All authors contributed to the study conception, design and drafting of the original manuscript. They reviewed, edited and approved the final manuscript.
	\section*{Appendix: Cardan's method}
	Let $q=u+v$, then we obtain 
	\beq
	q^3-3uvq-(u^3+v^3)=0. 
	\eeq 
	Now, comparing this equation with the cubic equation (\ref{cardan}) and we obtain
	\beq
	\ba{ll}
	uv=-b_1,u^3+v^3=-b_0,\\
	u^3=\frac{1}{2}\left(-b_0+\sqrt{b_0^2+4b_1^3}\right),\\ v^3=\frac{1}{2}\left(-b_0-\sqrt{b_0^2+4b_1^3}\right).
	\ea 
	\eeq
	\textbf{Case 1. $b_0^2+4b_1^3=\chi^2>0$}
	
	In this case $u^3, v^3$ are real numbers. Let $s$ denotes any root of $\frac{1}{2^{1/3}}\left(-b_0+\chi\right)^{1/3}$, then the values of $u$ are $s,\frac{(1+i\sqrt{3})}{2}s,\frac{(1-i\sqrt{3})}{2}s$ and the values of $v$ are defined by $-\frac{b_1}{s},-\frac{(1-i\sqrt{3})}{2}\frac{b_1}{s},-\frac{(1+i\sqrt{3})}{2}\frac{b_1}{s}$. 
	Then the values of $q$ are defined by 
	\beq
	\ba{ll}
	q_1 = s-\frac{b_1}{s},\\
	q_2 = \frac{(1 - i\sqrt{3}) s}{2}-\frac{(1 +i\sqrt{3})b_1}{2s} ,\\
	q_3 = \frac{(1 + i\sqrt{3}) s}{2}-\frac{(1 - i\sqrt{3}) b_1}{2s}. 
	\ea
	\eeq  
	Without loss of generality we can assume $s=\frac{sign(\chi-b_0)}{2^{1/3}}\left|\chi-b_0\right|^{1/3}$, where $sign(x)$ is the signum function gives the sign of $x$ and it is defined by $sign(x)=1$, if $x>0$ and $-1$ if $x<0$.

	\noindent\textbf{Case 2. $b_0^2+4b_1^3=-\chi^2<0$}
	
	In this case 
	\beq
	u^3=\frac{1}{2}\left(-b_0+i\chi\right), v^3=\frac{1}{2}\left(-b_0-i\chi\right) 
	\eeq 
	are complex numbers. Then the values of $u$ are defined by 
	\beq
	u=\sqrt{-b_1}\left[\cos\left(\frac{\theta}{3}\right)+i\sin\left(\frac{\theta}{3}\right)\right],~\sqrt{-b_1}\left[\cos\left(\frac{2\pi+\theta}{3}\right)+i\sin\left(\frac{2\pi+\theta}{3}\right)\right],~\sqrt{-b_1}\left[\cos\left(\frac{4\pi+\theta}{3}\right)+i\sin\left(\frac{4\pi+\theta}{3}\right)\right], 
	\eeq
	and the corresponding $v$ values are defined by 
	\beq
	v=\sqrt{-b_1}\left[\cos\left(\frac{\theta}{3}\right)-i\sin\left(\frac{\theta}{3}\right)\right],~\sqrt{-b_1}\left[\cos\left(\frac{2\pi+\theta}{3}\right)-i\sin\left(\frac{2\pi+\theta}{3}\right)\right],~\sqrt{-b_1}\left[\cos\left(\frac{4\pi+\theta}{3}\right)-i\sin\left(\frac{4\pi+\theta}{3}\right)\right],
	\eeq 
	where $-\pi<\theta=\tan^{-1}\left(-\frac{\chi}{b_0}\right)\le\pi$.
	Finally, the values of $q$ are defined by
	\beq
	\ba{ll}
	q_1 = 2\sqrt{-b_1}\cos\left(\frac{\theta}{3}\right),\\
	q_2 = 2\sqrt{-b_1}\cos\left(\frac{2\pi+\theta}{3}\right),\\
	q_3 = 2\sqrt{-b_1}\cos\left(\frac{4\pi+\theta}{3}\right). 
	\ea
	\eeq
	
	\noindent\textbf{Case 3. $b_0^2+4b_1^3=0$} 
	
	In this case
	\beq
	u^3=v^3=-\frac{b_0}{2}. 
	\eeq 
	are real and the values of $u$ and $v$ are defined by
	\beq
	\ba{ll}
	u=r,\frac{(1+i\sqrt{3})}{2}r,\frac{(1-i\sqrt{3})}{2}r,\\
	v=-\frac{b_1}{r},\frac{(1-i\sqrt{3})}{2}\frac{b_1}{r},\frac{(1+i\sqrt{3})}{2}\frac{b_1}{r}.
	\ea
	\eeq   
	where $r=\ds sig(-b_0)\left|-\frac{b_0}{2}\right|^{\frac{1}{3}}$. Therefore, the values of $q$ are defined by
	\beq
	\ba{ll}
	q_1 = r-\frac{b_1}{r},\\
	q_2 = \frac{(1 - i\sqrt{3}) r}{2}-\frac{(1 +i\sqrt{3})b_1}{2r} ,\\
	q_3 = \frac{(1 + i\sqrt{3}) r}{2}-\frac{(1 - i\sqrt{3}) b_1}{2r}. 
	\ea
	\eeq    
	
\end{document}